\documentclass[]{interact}

\usepackage{epstopdf}
\usepackage{subfigure}

\usepackage{natbib}
\bibpunct[, ]{(}{)}{;}{a}{}{,}

\theoremstyle{plain}

\theoremstyle{definition}

\theoremstyle{remark}

\usepackage{color}

\usepackage{amssymb}
\usepackage{amsmath}

\usepackage{mathtools}

\usepackage{graphicx}
\graphicspath{{images/}}

\newcommand*\diff{\mathop{}\!\mathrm{d}}

\usepackage{tikz}
\usepackage{pgfplots}

\pgfplotsset{
	node near coord/.style args={#1/#2/#3}{
		nodes near coords*={
			\ifnum\coordindex=#1 #2\fi
		},
		scatter/@pre marker code/.append code={
			\ifnum\coordindex=#1 \pgfplotsset{every node near coord/.append style=#3}\fi
		}
	},
	nodes near some coords/.style={ 
		font=\tiny,
		scatter/@pre marker code/.code={},
		scatter/@post marker code/.code={},%
		node near coord/.list={#1} 
	}
}

\definecolor{db}{rgb}{0.035, 0.239, 0.498} 

\definecolor{sphere1}	{rgb}{0.874, 0.749, 0.498} 
\definecolor{tube1}		{rgb}{0.000, 0.498, 0.247} 
\definecolor{surface}	{rgb}{0.000, 0.498, 1.000} 
\definecolor{tube0}		{rgb}{0.000, 0.749, 0.498} 
\definecolor{sphere0}	{rgb}{1.000, 1.000, 0.000} 

\newtheorem{conj}{Conjecture}

\begin{document}
	
\title{Experimental visually-guided investigation of sub-structures\\ in three-dimensional Turing-like patterns}

\author{
	\name{Martin Skrodzki\textsuperscript{a,b}\thanks{CONTACT M. Skrodzki. Email: mail@ms-math-computer.science}, Ulrich Reitebuch\textsuperscript{c}, and Eric Zimmermann\textsuperscript{c}}
	\affil{
		\textsuperscript{a}ICERM, Brown University, Providence, RI, USA\\
		\textsuperscript{b}RIKEN iTHEMS, Wako, Saitama, Japan\\
		\textsuperscript{c}Institut f\"{u}r Mathematik und Informatik, Freie Universit\"{a}t Berlin, Berlin, Germany
}
}

\maketitle

\begin{abstract}
		In his 1952 paper ``The chemical basis of morphogenesis'', Alan M.\@ Turing presented a model for the formation of skin patterns. 
		While it took several decades, the model has been validated by finding corresponding natural phenomena, e.g.\@ in the skin pattern formation of zebrafish. 
		More surprising, seemingly unrelated pattern formations can also be studied via the model, like e.g.\@ the formation of plant patches around termite hills. 
		In 1984, David A.\@ Young proposed a discretization of Turing's model, reducing it to an activator/inhibitor process on a discrete domain. 
		From this model, the concept of three-dimensional Turing-like patterns was derived.
		
		In this paper, we consider this generalization to pattern-formation in three-dimensional space.
		We are particularly interested in classifying the different arising sub-structures of the patterns.
		By providing examples for the different structures, we prove a conjecture regarding these structures within the setup of three-dimensional Turing-like pattern.
		Furthermore, we investigate--guided by visual experiments---how these sub-structures are distributed in the parameter space of the discrete model. 
		We found two-fold versions of zero- and one-dimensional sub-structures as well as two-dimensional sub-structures and use our experimental findings to formulate several conjectures for three-dimensional Turing-like patterns and higher-dimensional cases.
\end{abstract}


\begin{amscode}
	92C15, 
	37N25, 
	37B15, 
	92-05 
\end{amscode}
	
\section{Introduction}
\label{sec:Introduction}

Why do tigers and zebras have stripes while other animals like cows or leopards are spotted?
This question is one of the aspects of the broader task to understand \emph{morphogenesis}. 
Composed of the two Greek words \emph{morph{\'e}} (shape) and \emph{g\'enesis} (creation), the term describes the biological process of an organism developing its shape. 
One of the earliest authors to investigate this question was d'Arcy Wentworth Thompson, who devoted his 1917 treatise ``On Growth and Form'' to it, see~\cite{thompson1917growth}.
Another notable contribution was made by Alan M. Turing in his article ``The chemical basis of morphogenesis'', issued in the philosophical transactions of the royal society of London, see~\cite{turing1990chemical}.
While Thompson's approach to morphogenesis is largely based on different growth rates of the animal, Turing focuses---as suggested by the title---on a chemical mechanism giving rise to different skin color patterns.
He states in the abstract of his article ``that a system of chemical substances, called morphogens, reacting together and diffusing through a tissue, is adequate to account for the main phenomena of morphogenesis,'' \cite[p.~37]{turing1990chemical}.
We will give a detailed presentation of Turing's model in Section~\ref{sec:ChemicalModelTuring}.

Indeed, though much later, the patterns predicted by Turing have been found in biological settings and physical systems.
While they indeed describe certain animal skin patterns, they surprisingly also arise in larger biological phenomena, like the formation of termite hills.
Section~\ref{sec:TuringPatternsInBiologyAndRelatedApplications} lists several of these applications.

With the growing availability of computers, discretizations of models become increasingly important.
Regarding the concept of Turing patterns, a most notably contribution was made by David A. Young in his 1984 paper ``A local activator-inhibitor model of vertebrate skin patterns,'' see~\cite{young1984local}.
His model is not only discrete, but also reduces Turing's setup to two simple morphogens with clear functionalities: one activator morphogen that causes cells to be differentiated (colored) and one inhibitor morphogen that prevents the differentiation of cells.
The structures obtained from Young's discretizations of Turing's work are consequentially referred to as Turing-like patterns.
While Young does never state it in his paper, he basically gave the description of a two-dimensional cellular automaton to produce the patterns. 
We present Young's approach in detail in Section~\ref{sec:YoungsActivatorInhibitorDiscretization} and give some general introduction to cellular automata in Section~\ref{sec:CellularAutomata}.

The formulation of Turing-like patterns by Young allows for a very efficient evaluation of different parameters and setups.
While Turing has ``obtained [his results] in a few hours by a manual computation''~\cite[p.~59]{turing1990chemical}, modern computers can create respective imagery within seconds.
This makes these patterns interesting for different applications.
For instance, a series of articles has been devoted to generalizations of two-dimensional patterns to three-dimensional structures, see~\cite{skrodzki2016chladni,skrodzki2017turing,skrodzki2018mondrian}.
The second paper considered the emergence of Turing-like patterns from a corresponding three-dimensional cellular automaton, see Section~\ref{sec:SteppingUpOneDimension} for a detailed description.
The resulting three-dimensional patterns in particular pose a visualization challenge: It is no longer feasible to render each cell as a colored solid as the outermost cells would obscure the view to the inside of the pattern.
To illustrate the obtained findings, only the borders between differentiated and undifferentiated cells were shown, see Figure~\ref{fig:Bridges}.
The authors state at the end of their paper: ``\textellipsis with varying parameters, the [two-dimensional] pattern performs a phase transition from disconnected points (`zero-dimensional') to connected stripes (`one-dimensional'). A similar behavior is exhibited by three-dimensional patterns\textellipsis. One would expect that in three-dimensional patterns also `two-dimensional' connections occur,''~\cite[p.~417]{skrodzki2017turing}.
However, they were unable to present a corresponding set of parameters to create such patterns.

This article is dedicated to closing this gap.
After discussing the works mentioned above in greater detail, we present:
\begin{itemize}
	\item a classification of lower-dimensional sub-structures in Turing-like patterns,
	\item an experimental investigation of these sub-structures in the three-dimensional case,
	\item examples for all possible sub-structures, i.e.\@ a proof to the conjecture raised by~\cite{skrodzki2017turing},
	\item and several conjectures on the sub-structures and parameter spaces for both the three-dimensional case and higher-dimensional cases.
\end{itemize}

Special emphasis is put on understanding and visualizing not only the Turing-like patterns, but also their underlying parameter space.
As we do not have a thorough mathematical description available to formally distinguish different sub-structures in Turing-like patterns, see Section~\ref{sec:PartitioningTheParameterSpaceIn3D}, our approach is experimental and visually guided.
This is possible, as the different sub-structures are easily distinguishable by a human observer.
Therefore, we can discretize the parameter space and classify each element by visual inspection.
As the number of elements is very large, we perform this classification task within a citizen science project, see Section~\ref{sec:TheExperiment}.
This approach provides us with a labeled partitioning of the parameter space.
By visual and statistical analysis of these results, we are able to formulate conjectures for both the continuous setting as well as higher-dimensional analogs.

\section{Related Work}
\label{sec:RelatedWork}

To set the stage for our work, we briefly present related research from four different areas.
First, we consider how the morphogenetic patterns by Turing actually arise in biological contexts.
Second, we consider basic literature on cellular automata and how these mechanisms in turn inspire research in biology.
Third, as one of the main focuses of this article is on visualization and illustration, we consider how Turing(-like) patterns have given rise to different visual artworks.
Finally, as our experiment is concerned with the parameter space of three-dimensional Turing-like patterns, we present previous results on the analysis of parameter spaces of two-dimensional patterns as well as on three-dimensional patterns.

\subsection{Turing patterns in biology and related applications}
\label{sec:TuringPatternsInBiologyAndRelatedApplications}

\begin{figure}
	\centering
	\includegraphics[width=0.75\textwidth]{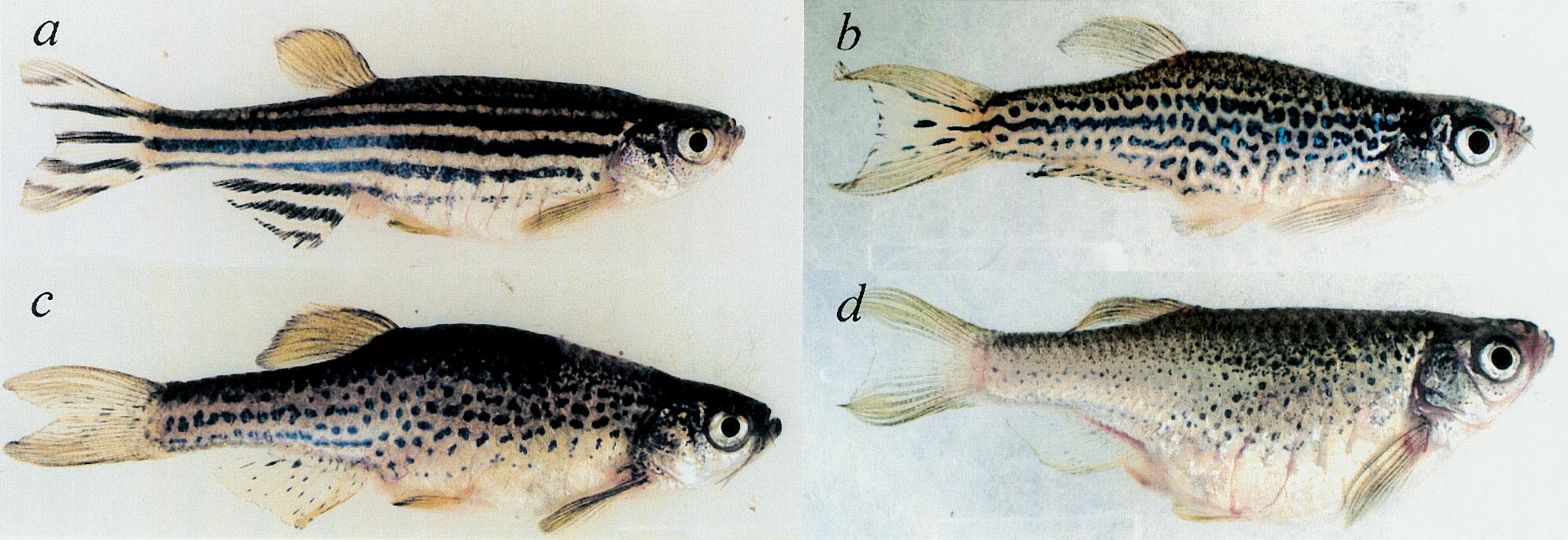}
	\caption{Four fully-grown (5–6 months) female zebrafish, with different alleles of the leopard gene, exhibiting striped and dappled patterns respectively. Image taken from~\cite{asai1999zebrafish}.}
	\label{fig:Zebrafish}
\end{figure}

While Turing predicted his morphogenetic patterns in the 1950s, it was only several decades later that these patterns were actually discovered in animal skins.
This long process gave Turing patterns a bad standing in the community, which was alleviated with more and more examples for self-regulated, reaction-diffusion based pattern formation in nature, see~\cite{kondo2010reaction}.
A particular, well-studied example in this context is the zebrafish, see Figure~\ref{fig:Zebrafish}.
Studies have connected genes~(\cite{asai1999zebrafish}), chemical properties~(\cite{wertheim2019can}), and cell-network interactions~(\cite{nakamasu2009interactions}) to the formation of Turing patterns within the zebrafish.
Recently, researchers revealed how mechanical stresses can lead to tissue anisotropies and thus morphogen gradients in the formation of Turing patterns, causing the characteristic stripes of a common zebrafish, see~\cite{hiscock2015orientation}.
Other works also incorporate different patterns to combine e.g.\@ stripe and dot formation to show how more complex structures can emerge from Turing's simple basics, see for instance~\cite{scoones2020dot}.

While Turing patterns thus describe actual skin coloring of certain animals in nature, they have further, surprising applications in the realm of biology.
The reaction-diffusion theory started by Turing forms the mathematical foundation of the notion of scale-dependent feedback (SDF).
These mechanisms play an important role in the spatial self-organization of ecosystems.
Recent works provide evidence that vegetation patterns generated by ants, termites, and other subterranean animals are inherently caused by such feedback procedures, see~\cite{pringle2017spatial}.
The authors also provide a list of references which cover the formation of gaps, labyrinths, stripes,
spots, and rings in different ecosystems, such as tropical deserts and savannas as well as subarctic peatlands, subtropical swamps, montane forests, Pacific reefs, and Atlantic mussel beds.
On a way smaller scale, diffusion-activation patterns help to understand processes in the visual cortex.
In this context, they can be used for instance to model geometric visual hallucinations, see~\cite{bressloff2002geometric}.

\subsection{Cellular automata}
\label{sec:CellularAutomata}

As hinted at in Section~\ref{sec:Introduction} and as will be explained in detail in Section~\ref{sec:YoungsActivatorInhibitorDiscretization}, the continuous formulation of Turing was discretized in the form of a cellular automaton (CA) by~\cite{young1984local}.
First steps towards CA have been made by Stanislav Ulam and John von Neumann in the late 1940s~\cite[cf.][p.~876]{wolfram2002new}.
However, the mechanism of CA only became popular with John H. Conway's \emph{Game of Life} in the 1970s, reported on in the article by~\cite{gardener1970mathematical}.
After several articles on cellular automata in the 1980s, Stephen Wolfram published his widely received treatise ``A new kind of science'', see~\cite{wolfram2002new}.
Despite criticism on several parts of the book as being wrong or unscientific, it serves as a concise introduction to the theory and applications of CA.

Generally, a CA is given by:
\begin{itemize}
	\item a discrete domain of cells, 
	\item a finite neighborhood relation between the cells, 
	\item a set of states each cell can be in, 
	\item and a transition function which attributes a new state to each cell, based on the cell states in its neighborhood. 
\end{itemize}
In the case of Turing-like patterns as discretized by Young, there are only two different states.
Namely, cells can be either differentiated or not. 
The neighborhood relation is based on a Moore neighborhood, also referred to as the $1$-ring under the Chebyshev distance.
Therefore, in a two-dimensional square grid, it consists of the eight cells around a center cell that share at least one corner with the center cell.
Accordingly, the three-dimensional Moore neighborhood includes $26$ cells.
The transition function will be described in detail in Section~\ref{sec:YoungsActivatorInhibitorDiscretization}.

Interestingly, CA are themselves the object of study in the search for a basic understanding of life.
An article by~\cite{chan2018lenia} explores the emergence of life-like structures from CA.
Relevant to the work in this paper, the author also provides descriptions of the parameter spaces for the described systems.
Several different behaviors are distinguished both in a two-dimensional~\cite[Fig.~9]{chan2018lenia} and a three-dimensional~\cite[Fig.~10]{chan2018lenia} parameter space.
The latter is related to our investigations in Section~\ref{sec:PartitioningTheParameterSpaceIn3D}.

In a series of articles,~\cite{beer2004autopoiesis,beer2014cognitive,beer2015characterizing,beer2020investigation} uses elements from the \emph{Game of Life} both as model and study object.
The author applies the notion of \emph{autopoiesis} to the comparably compact setup of the cellular automaton to investigate its consequences, implications, and shortcomings.
While these works tackle large-scale questions in the emergence of life and cognition, the work of~\cite{tonello2019boolean} considers the small-scale question of inhibition-models in the context of cell-to-cell communication.
Similar to the activator-inhibitor approach presented in Section~\ref{sec:YoungsActivatorInhibitorDiscretization}, their model assumes differentiated and undifferentiated cells, characterized by \emph{delta} and \emph{notch}.
Similar to the work of Beer, they also consider the stability of patterns under perturbations.
We will have to keep these \emph{basins} of stability in mind when interpreting the partitioning of the parameter space in our application.

\subsection{Turing(-like) patterns in illustration and arts}

Because of their simple formulation as well as their relation to naturally occurring patterns, Turing patterns have lent themselves to quite some extent to illustration and arts.
By rendering a cell based on the variation around it,~\cite{mccabe2010cyclic} arrives at the notion of multi-scale Turing patterns.
By imposing symmetry conditions, the author creates a juxtaposition of both mathematically complex and organic looking imagery.
In order to speed up the generation of these patterns,~\cite{schwehm2016fast} introduces a GPU-based algorithm.
The reached performance allows for real-time interaction with the system, as was done in a corresponding art installation.
Furthermore, the usage of color is explored within the renderings.
Turning not to GPUs, but to the Photoshop software,~\cite{werth2015turing} creates Turing patterns by iteratively blurring and sharpening an initial white noise image.
The author uses these patterns as the starting point in the design of abstract acrylic paintings.

While these works employ Turing patterns, the approach of~\cite{greenfield2016turing} uses Young's discretization scheme.
The author experiments with non-uniform distributions for cells being initially differentiated and also provides examples for non-circular areas of influence for both the activator and inhibitor.
Final results are achieved by overlaying several different images.
This cursory overview on Turing(-like) patterns in the context of illustration and arts proves their versatility as well as their wide usage within this field.
The work of~\cite{greenfield2016turing} served as a motivation for~\cite{skrodzki2017turing} which in turn motivates the research presented in this article.

\subsection{Parameter spaces and three-dimensional Turing patterns}

Several works have been devoted to understanding the parameter space of two-dimensional Turing(-like) patterns.
In the context of a predator-prey model,~\cite{weide2020pattern} consider the spatial distribution of prey and find several qualitative transitions when changing their model parameters.
Namely, they transition from spots of undifferentiated cells, via differentiated stripes and undifferentiated stripes to spots of differentiated cells~\cite[Fig.~5]{weide2020pattern}.
Taking a step further, considering the variation of two parameters results in two-dimensional maps indicating a wide range of behaviors, from all-undifferentiated cells via the states listed above to all-differentiated cells~\cite[cf.][Fig.~8,9]{ishida2020emergence}.
Similar results can be found in the investigation of dynamic Ising models, see~\cite{merle2019turing}.
Note that these investigations of the parameter space alter one or two parameters simultaneously.
We divert from this, by identifying a set of three parameters and by investigating an entire section  of the three-dimensional parameter space.

Furthermore, we are interested in the parameter space of three-dimensional rather than two-dimensional Turing-like patterns.
In the context of diffusion-reaction in microemulsion, researchers assert that ``actual [Turing] patterns can never be truly planar\textellipsis at a molecular level,''~\cite[p.~1309]{bansagi2011tomography}.
The authors consider the Belousov-Zhabotinsky reaction within a cylindrical domain and provide experimental evidence for zero-dimensional ``spot'' patterns, one-dimensional ``labyrinthine'' patterns, and two-dimensional ``lamellar'' patterns.
While they provide experiments and numerical simulations for each of these cases, it remains unclear how much the shape of the domain effects the pattern formation.
Furthermore, it is unclear what parameter changes cause transitions from one pattern to another or whether the effect scales to larger domains.

When going to a cubical domain with periodic boundary conditions, similar patterns as listed above can be observed, see~\cite{leppanen2002new}.
The authors state that ``starting from completely random initial conditions, it is not likely for the
system to converge into a purely lamellar state,''~\cite[p.~38]{leppanen2002new}.
Indeed, the authors obtain two-dimensional patterns by manually choosing the initially differentiated cell sets.
Again, it remains unclear how stable the respective patterns are, i.e.\@ how slight parameter changes effect the pattern formation and how the patterns scale with the domain size.
These shortcomings of previous work motivate us to investigate the complete parameter space of three-dimensional Turing-like patterns.

\section{Turing patterns and their Turing-like discretizations}
\label{sec:TuringPatternsAndDiscretizations}

In this section, we will present the underlying theoretical models that form the basis of our experiments.
First, we will describe Turing's chemical model of morphogenesis.
Second, we present the discretization into an activator-inhibitor model by Young.
Finally, we show how Young's model can be transferred into a three-dimensional setup.
For this, we briefly discuss the resulting parameter space as well as challenges to the visualization of the corresponding patterns.

\subsection{The chemical model of Turing}
\label{sec:ChemicalModelTuring}

As stated in Section~\ref{sec:Introduction}, the morphogenetic model of Turing is based on a reaction-diffusion system of chemical substances.
While his model allows arbitrary numbers of morphogenetic substances, throughout most of his article, Turing assumes two morphogens,~$X$ and~$Y$, which are produced at a constant rate.
Furthermore, the following interactions between the two morphogens are established:~$Y$ is destroyed depending on the concentration of~$Y$,~$X$ is converted into~$Y$  at a rate depending on~$X$, $X$ is additionally produced depending on the concentration of~$X$, and~$X$ is destroyed at a rate depending on~$Y$.

More specifically, Turing assumes a ring of~${N\in\mathbb{N}}$ cells, with the two morphogens~$X$ and~$Y$ present in cell~${r=1,\ldots,N}$ via concentrations~$X_r$ and~$Y_r$ respectively.
Furthermore, he assumes cell-to-cell diffusion of~$X$ by~$\mu$ and of~$Y$ by~$\nu$.
Finally, the last assumption is that the concentrations of~$X$ and~$Y$ are increasing:~$X$ at the rate~$f(X,Y)$ and~$Y$ at the rate $g(X,Y)$.
Given this setup, the behavior of the system may be described by the~$2N$ differential equations
\begin{equation*}
	\begin{drcases}
		\frac{\diff X_r}{\diff t}=f(X_r,Y_r)+\mu(X_{r+1}-2X_r+X_{r-1})\\
		\frac{\diff Y_r}{\diff t}=g(X_r,Y_r)+\nu(Y_{r+1}-2Y_r+Y_{r-1})
	\end{drcases}
	r=1,\ldots,N.
\end{equation*}
In the case that the functions~$f(X,Y)$ and~$g(X,Y)$ can be approximated linearly, Turing provides general solutions of the above set of differential equations~\cite[cf.][Sec.~6]{turing1990chemical}. Furthermore, he extends this discrete model from a set of~$N$ cells to a continuous ring of tissue~\cite[cf.][Sec.~7,8]{turing1990chemical}.

\begin{figure}
	\centering
	\includegraphics[width=0.5\textwidth]{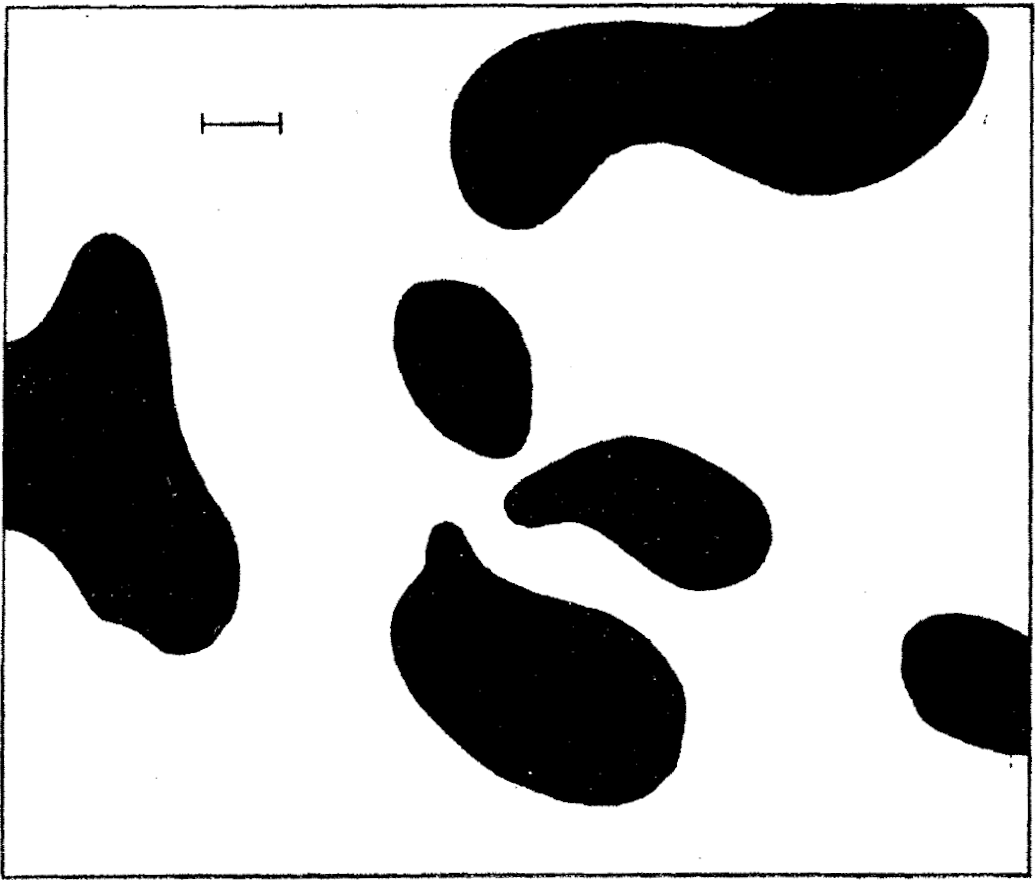}
	\caption{A two-dimensional result, obtained by Turing from his morphogen system~\cite[Fig.~2]{turing1990chemical}. Here, a one-morphogen system is used, the line indicates a marker of unit length.}
	\label{fig:Turing}
\end{figure}

To give a two-dimensional example, Turing turns to a homogeneous one-morphogen system. To this system, he applies random disturbances without diffusion for a period of time and then diffusion without disturbance. Namely, he chooses numbers~$u_{r,s}$ initially at random to be~$\pm1$ with equal probability. Then, he considers the function
\begin{equation*}
	f(x,y)=\sum_{r,s} u_{r,s}\exp\left(-\frac{1}{2}\left((x-hr)^2+(y-hs)^2\right)\right),
\end{equation*}
where~$h\approx0.7$. After applying diffusion and perturbation as indicated above, he obtains a figure, where a point~$(x,y)$ is colored black if~$f(x,y)$ is positive, see Figure~\ref{fig:Turing}~\cite[cf.][Sec.~9]{turing1990chemical}.

Interpreting his figure, Turing states: ``This is the least interesting of the cases. It is possible, however, that it might account for `dappled' colour patterns[\textellipsis]''~\cite[p.~66]{turing1990chemical}.
Later on, he continues to interpret his results within the context of actual biological processes. Namely, he reasons: ``If dappled patterns are to be explained in this way they must be laid down in a latent form when the foetus is only a few inches long. Later the distances would be greater than the morphogens could travel by diffusion.''~\cite[p.~66--67]{turing1990chemical}.
We have seen in Section~\ref{sec:TuringPatternsInBiologyAndRelatedApplications} that Turing's model actually proved to be correct for several developments of skin patterns in the animal kingdom.
See~\cite{woolley2017turing} for a more detailed discussion on both the theoretical background and the computation of Turing patterns on arbitrary surfaces.
However, Turing himself recognized that the used equations can only be approximately computed by hand, in a tedious process. 
Thus, subsequent work was devoted to providing simpler computational models for Turing's patterns.

\subsection{Young's activator-inhibitor discretization}
\label{sec:YoungsActivatorInhibitorDiscretization}

Several years after Turing presented his model, it was still not tested experimentally, but evidence suggested even more localized inter-cellular interaction than predicted by the model. 
Based on previous models for patterns in the visual cortex of the brain~(\cite{swindale1980model}), Young presented his own two-morphogen model, see~\cite{young1984local}.
He assumes two types of cells: differentiated (pigmented) cells (DCs) and undifferentiated cells (UCs).
Initially, the cell type is uniformly distributed in the domain.
Then, each DC produces an activator morphogen~$M^1$ which differentiates nearby UCs and an inhibitor morphogen~$M^2$ which causes nearby DCs to become undifferentiated. 
Both morphogens diffuse through the domain, with the inhibitor morphogen having longer range.
The UCs are passive in this model and do not produce any morphogen.
The diffusion of the morphogens is modeled as
\begin{equation*}
	\frac{\diff M^i}{\diff t}=\nabla\cdot D\cdot\nabla M^i-KM^i+Q,
\end{equation*}
where~${M^i = M^i(r, t)}$,~${i=1,2}$ is the morphogen (either activator or inhibitor) concentration at location~$r$ and time~$t$, while the other terms describe diffusion, chemical transformation, and production of the respective morphogen, see~\cite{young1984local}.

\begin{figure}
	\centering
	\includegraphics[width=0.75\textwidth]{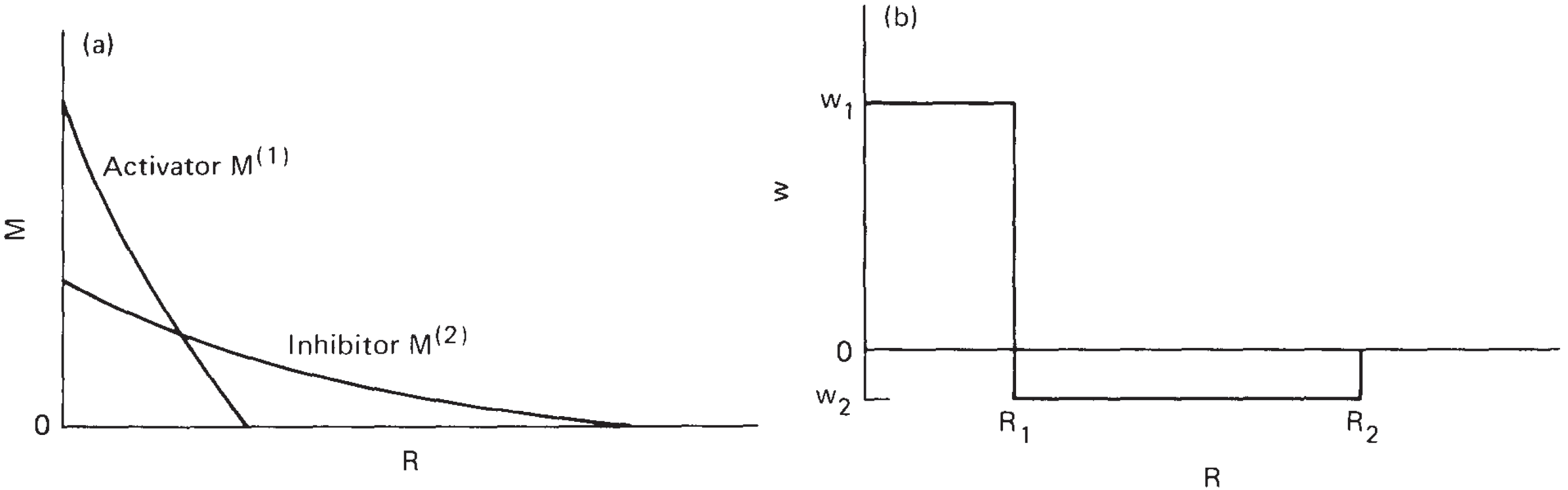}
	\caption{Left: Distribution field around a discriminated cell, given by the continuous diffusion of activator~$M^1$ and inhibitor~$M^2$ according to radius~$R$ around the cell.	Right: Reinterpreting the impact of the two morphogens around a discriminated cell by discretizing their summed contribution to an area of positive or negative influence, depending on the radius~$R$ around the cell.~\cite[Fig.~1]{young1984local}}
	\label{fig:Young_1}
\end{figure}

With time passing, each DC produces the two morphogens at a constant rate and both diffuse according to the above equation. 
This causes a distribution field around each DC, see Figure~\ref{fig:Young_1} (left).
Combining the two morphogens, attributing a positive value to the activator and a negative value to the inhibitor, the field has a positive as well as a negative region, depending on the distance~$R$ to the considered DC, see Figure~\ref{fig:Young_1} (right).
Here, Young considers a field of constant positive circular region of radius~$R_1$, surrounded by an annulus of constant negative influence of inner radius~$R_1$ and outer radius~$R_2$.

\begin{figure}
	\centering
	\includegraphics[width=0.5\textwidth]{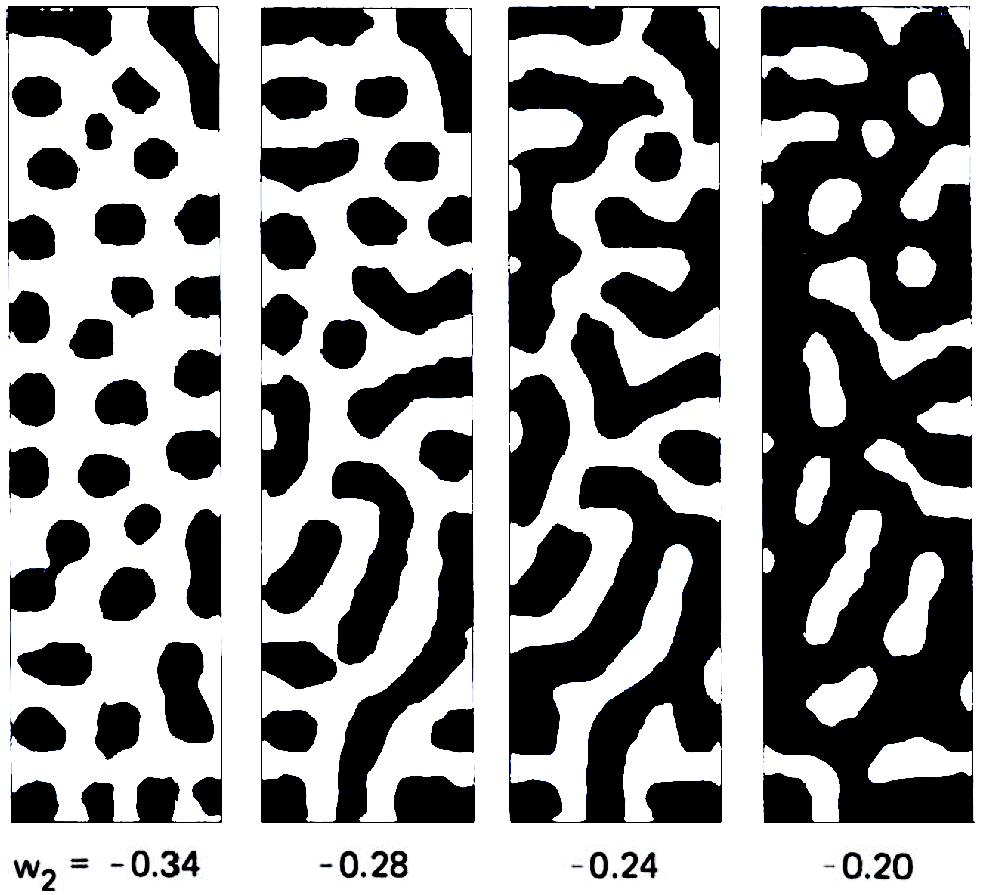}
	\caption{A Turing-like pattern, created by Young on~${25\times100}$ grid cells, with~${R_1=2.3}$, ${R_2=6.01}$, ${w_1=1.0}$, and~$w_2$ as indicated below the patterns~\cite[Fig.~2]{young1984local}.}
	\label{fig:Young_2}
\end{figure}

From this model, Turing-like patterns are obtained by starting with a rectangular grid of cells.
We will model a cell at position~$(x,y)$ in the grid by its \emph{state}~$s_t(x,y)$ at time~$t$, given by
\begin{equation*}
	s_t(x,y):=\begin{cases}
		1 & (x,y)\text{ is a DC},\\
		0 & (x,y)\text{ is a UC}.
	\end{cases}
\end{equation*}
Initially, DCs are randomly chosen in the grid, i.e.~${s_0(x,y)=1}$ for some cells chosen by some random process.
Now, to determine the state of a cell~$(x,y)$ in the next time step~$t+1$, we compute
\begin{equation}
\label{equ:TuringLikeIteration}
	s_{t+1}(x,y)=\begin{cases}
		1 			& \sum_{(x',y')\in B_{R_2}(x,y)}\omega_{t,(x,y)}(x',y') > 0,\\
		s_t(x,y) 	& \sum_{(x',y')\in B_{R_2}(x,y)}\omega_{t,(x,y)}(x',y') = 0,\\
		0 			& \sum_{(x',y')\in B_{R_2}(x,y)}\omega_{t,(x,y)}(x',y') < 0,\\
	\end{cases}
\end{equation}
where $B_{R_2}(x,y)$ is the ball of radius $R_2$ around $(x,y)$ and
\begin{equation}
\label{equ:TuringLikeWeighting}
	\omega_{t,(x,y)}(x',y') = \begin{cases}
		0 					& (x-x')^2+(y-y')^2 > R_2^2,\\
		w_1\cdot s_t(x',y') & (x-x')^2+(y-y')^2 < R_1^2,\\
		w_2\cdot s_t(x',y') & \text{otherwise}.
	\end{cases}
\end{equation}
That is, for any grid cell~$(x,y)$, all DCs within the circular region of radius~$R_2$ are taken into account. 
Those that lie within the smaller circular region of radius~$R_1$ contribute weight~$w_1$ while those in the annulus between~$R_1$ and~$R_2$ contribute weight~$w_2$. 
If the sum of these weights is positive, the cell becomes differentiated; if the sum is zero, the cell does not change its state; if the sum is negative, the cell becomes undifferentiated.
Young presents a corresponding experiment on~${25\times100}$ grid cells, with~${R_1=2.3}$, ${R_2=6.01}$, $w_1=1.0$, and~$w_2$ as indicated below the patterns, see Figure~\ref{fig:Young_2}.
He does not report his boundary conditions, i.e.\@ whether his grid is following a flat, a cylindrical, or a toric topology.

In his paper, Young proceeds to present a second, elliptical kernel for his patterns.
This anisotropic kernel---expectably---creates stripe patterns oriented along the longer axis of the ellipse~\cite[see][Fig.~3]{young1984local}.
Without having mentioned it, he reinterpreted Turning's model in the context of cellular automata (CA).
All images presented so far where two-dimensional, but Young's CA offers an easy generalization to arbitrary dimensions.

\subsection{Stepping up one dimension}
\label{sec:SteppingUpOneDimension}

Turing's model as described in Section~\ref{sec:ChemicalModelTuring} has been applied to three-dimensional settings in the medical~(\cite{bansagi2011tomography}) and physical~(\cite{leppanen2002new}) realm. Analogously, the discretization of Young also gives rise to three-dimensional Turing-like patterns.
The single necessary addition is a term~${(z-z')^2}$ in the right-hand side of Equation~(\ref{equ:TuringLikeWeighting}).
To further simplify Young's approach, it can be reduced to choosing initial DCs in the domain uniformly randomly with some probability~${\rho\in[0,1]}$ and furthermore fixing~${w_1=1}$, ${w_2=-1}$, see~\cite{skrodzki2017turing}.
Given a domain size, these simplifications reduce the parameter space of Turing-like patterns to choosing a probability~${\rho\in[0,1]}$ as well as two radii~$R_1$ and~$R_2$.

\begin{figure}
	\centering
	\includegraphics[width=0.49\textwidth]{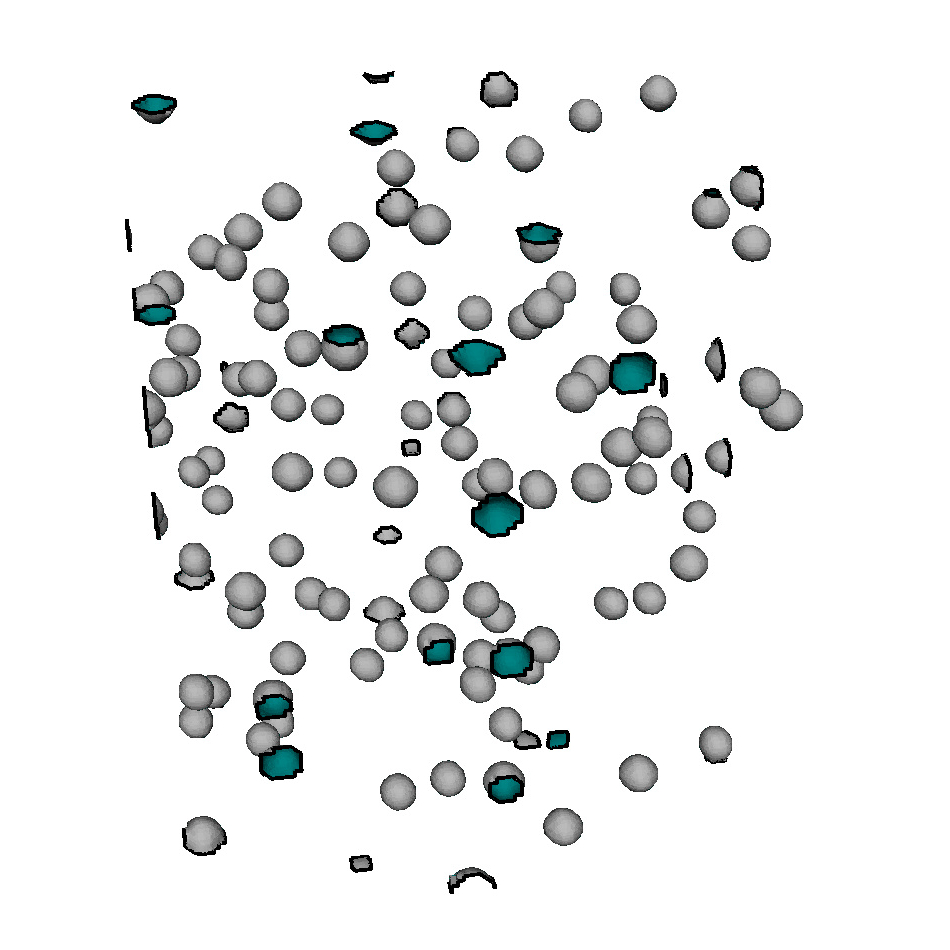}
	\hfill
	\includegraphics[width=0.49\textwidth]{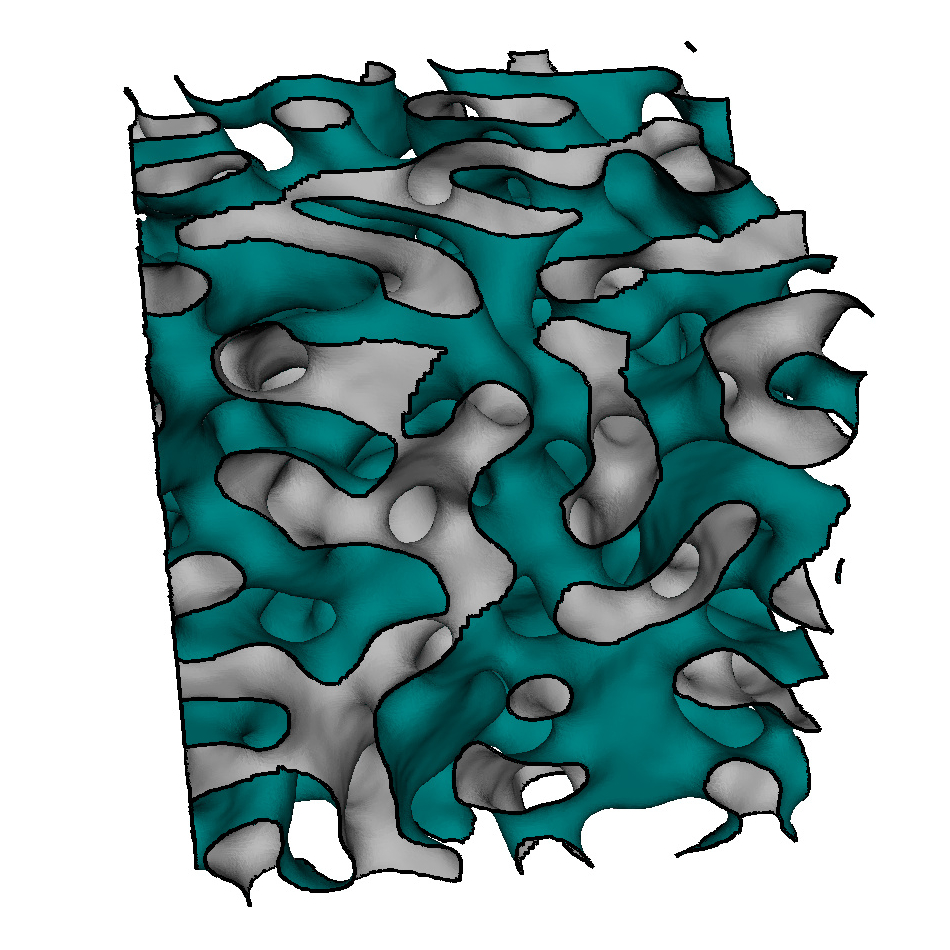}
	\caption{Two Turing-like patterns on a three-torus, discretized into~${100\times100\times100}$. Left:~${\rho=0.9999,R_1=4,R_2=7}$. Right:~${\rho=0.5,R_1=8,R_2=10}$.~\cite[Fig.~4]{skrodzki2017turing}}
	\label{fig:Bridges}
\end{figure}

In the two-dimensional patterns of Figures~\ref{fig:Turing} and~\ref{fig:Young_2}, DCs are colored black, while UCs are rendered white. 
For the visualization of three-dimensional patterns, a different approach is necessary in order to gain insights into the presented structures. 
Thus, the images from~\cite{skrodzki2017turing} show only the boundary between regions of DCs and UCs. 
Furthermore, these images are rendered on the domain of a three-torus, i.e.\@ two opposing sides of the cube are identified.
Examples for three-dimensional Turing-like patterns on a domain of~${100\times100\times100}$ cells in a three-torus, taken from~\cite{skrodzki2017turing}, are given in Figure~\ref{fig:Bridges}.

The publication by~\cite{skrodzki2017turing} on three-dimensional Turing-like patterns ends with the following paragraph: ``Note how\textellipsis with varying parameters, the pattern performs a phase transition from disconnected points (`zero-dimensional') to connected stripes (`one-dimensional'). A similar behavior is exhibited by three-dimensional patterns\textellipsis. One would expect that\textellipsis also `two-dimensional' connections occur.''~\cite[p.~418]{skrodzki2017turing}. The remainder of this paper is devoted to systematically exploring the parameter space of three-dimensional Turing-like patterns and to investigate the variety of obtainable structures.

\section{Partitioning the parameter space in 3D}
\label{sec:PartitioningTheParameterSpaceIn3D}

In order to experimentally investigate the parameter space, we first fixed the size of the cubical three-torus to work on.
To be able to fully classify the entire parameter space in a somewhat interactive fashion, we chose a grid size of~$ {70\times70\times70} $.
We base our parameter space on the discretizations of~\cite{young1984local} and the weighting choices~$\omega_1=1$,~$\omega_2=-1$ of~\cite{skrodzki2017turing}, as discussed in the above Section~\ref{sec:SteppingUpOneDimension}.
Thus, the parameter space consists of three parameters: radius~$ {R_1 \in \mathbb{N}} $ of the activator morphogen, radius~$ {R_2 \in \mathbb{N}} $ of the inhibitor morphogen, and initial cell activation probability~$ {\rho \in [0,1]} $.

\begin{figure}
	\centering
	\begin{tikzpicture}[scale=0.7]
		\pgfplotsset{		
			samples=150,
			axis lines=left,
		}
		\begin{axis}[
		legend pos = outer north east,
		xlabel={$x$},
		ylabel={$\sigma(x)$},
		]
		\addplot[db, thick, domain=0:120] {exp(\x/5-12) / (1 + exp(\x/5-12))};
		\end{axis}
	\end{tikzpicture}
	\caption{Sigmoid for initial cell activation proposed in Equation~(\ref{equ:sigmoid}). The domain~$ {\mathcal{I}=\{0,\ldots,120\} }$ is mapped onto probability values~$ {\sigma(x)\in[0,1]} $.}
	\label{fig:sigmoidPlot}
\end{figure}
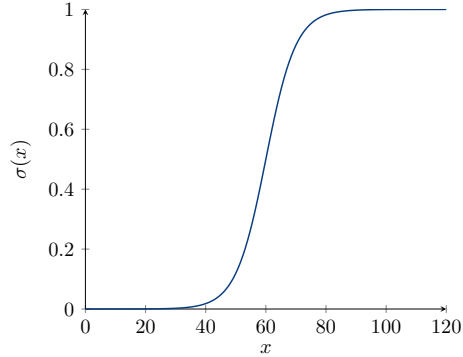

Choosing too large values for the radii~$R_1$ and~$R_2$ in a comparably small domain would obscure the patterns formed.
Therefore, we reduced the choice for the two radii to be from~$ {\{1,\ldots,40\}\subset\mathbb{N}} $.
The value~$\rho$ determines the probability of initial activity of each cell in our cubical grid.
To discretize the probability, we chose a domain of~$ {\mathcal{I}\coloneqq\{0,\ldots,n_\rho\}\subset\mathbb{N}} $ with~$ {n_\rho=120} $ in our experiment.
As the probabilities~$ {\rho=0} $ and~$ {\rho=1} $ only provide completely undifferentiated or completely differentiated domains respectively, we chose the lowest probability to be investigated to be at~$ {\exp(-\min_\rho)} $ with~$\min_\rho=12$, i.e.\@ the lowest probability as~$ {\exp(-12)\approx 6\cdot10^{-6}} $.
As~\cite{skrodzki2017turing} found their zero-dimensional behavior for extremely large values of~$\rho$, we decided to discretize this dimension of the parameter space non-linearly.
Therefore, the domain is mapped to a probability value via the following sigmoid
\begin{equation}
	\begin{array}{rcl}
		\sigma : \mathcal{I} & \to & [0,1], \\
		x & \mapsto & \frac{\exp\left(2 \cdot \min_\rho \cdot n_\rho^{-1} \cdot x - \min_\rho\right)}{1 + \exp\left(2 \cdot \min_\rho \cdot n_\rho^{-1} \cdot x - \min_\rho\right)},
	\end{array}
\label{equ:sigmoid}
\end{equation}
which provides good resolutions at very low and very high probabilities as well as a decent resolution of the behavior between these extremes, see Figure~\ref{fig:sigmoidPlot}.
In the following, we will assume that some~$ {x\in\mathcal{I}} $ was chosen and simply write~$ {\rho\coloneqq\sigma(x)} $ for brevity.

\begin{figure}
	\centering
	\subfigure[\colorbox{gray!10}{\textcolor{sphere1}{DC spheres $\blacksquare$}}]{
		\includegraphics[width=.17\textwidth]{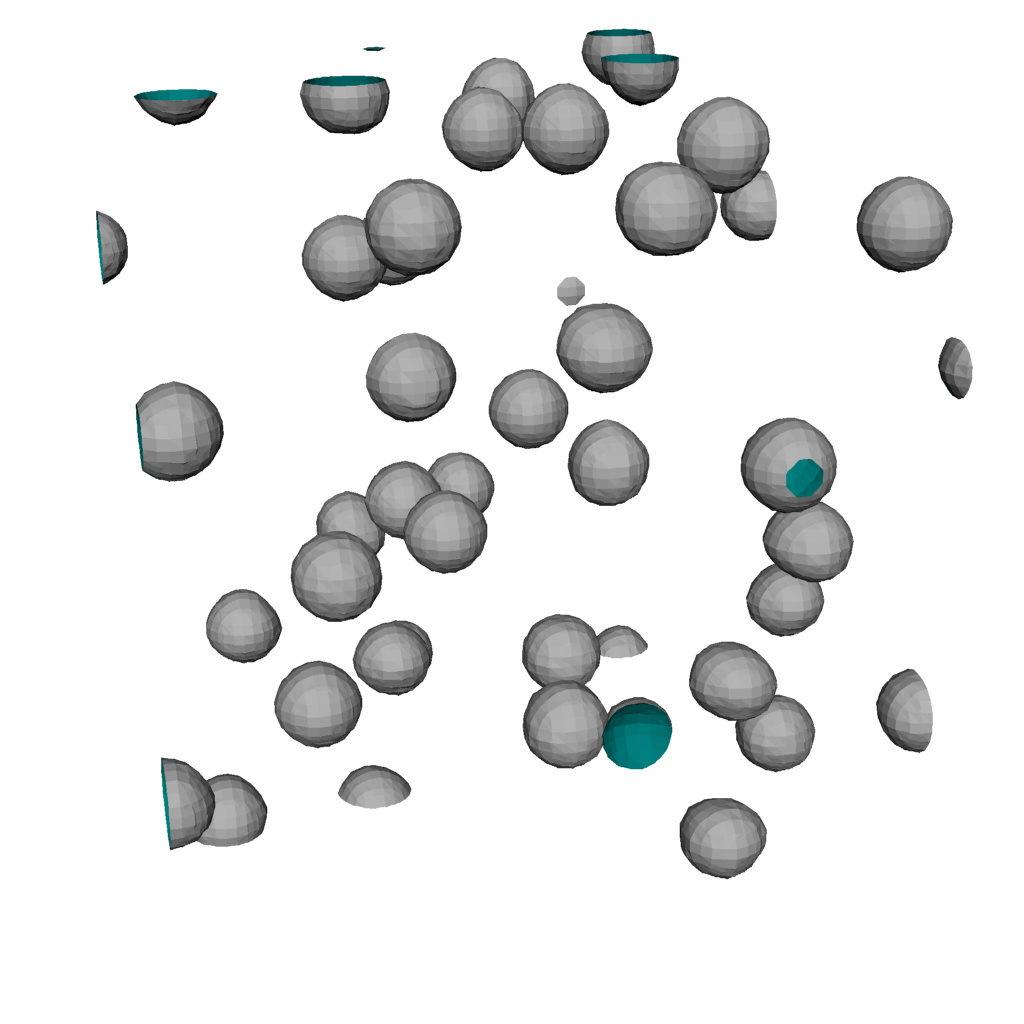}
		\label{fig:3DPatternsW0d}
	}
	\hfill
	\subfigure[\colorbox{gray!20}{\textcolor{tube1}{DC pipes $\blacksquare$}}]{
		\includegraphics[width=.17\textwidth]{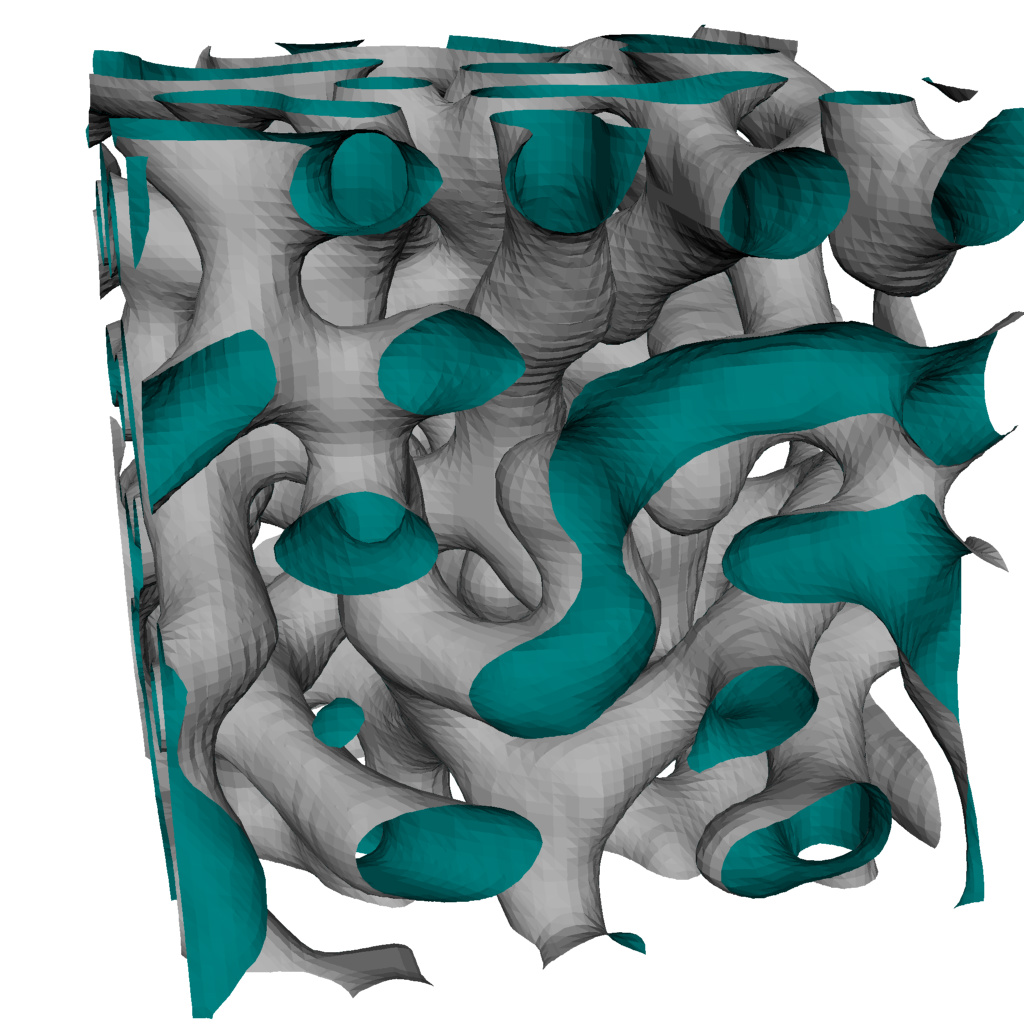}
		\label{fig:3DPatternsW1d}
	}
	\hfill
	\subfigure[\colorbox{gray!30}{\textcolor{surface}{areas $\blacksquare$}}]{
		\includegraphics[width=.17\textwidth]{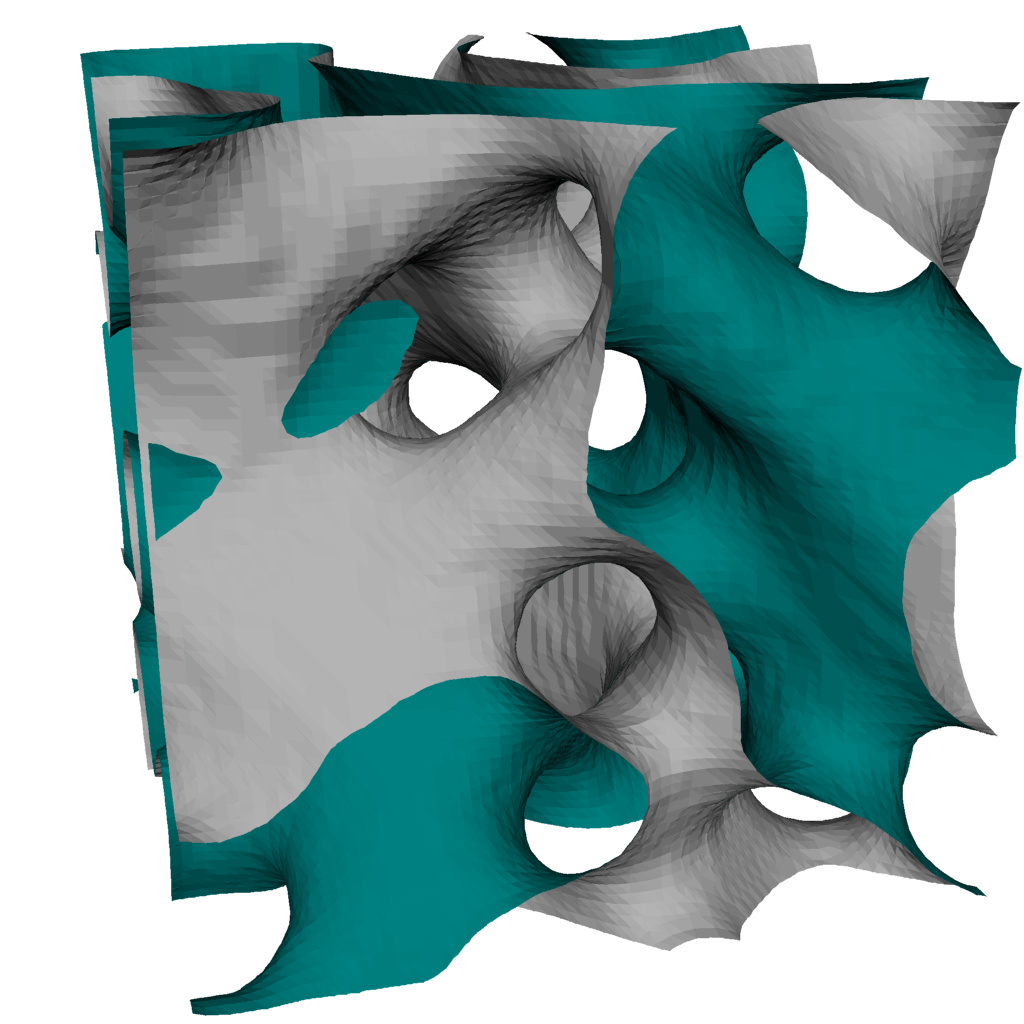}
		\label{fig:3DPatternsWB2d}
	}
	\hfill
	\subfigure[\colorbox{gray!40}{\textcolor{tube0}{UC pipes $\blacksquare$}}]{
		\includegraphics[width=.17\textwidth]{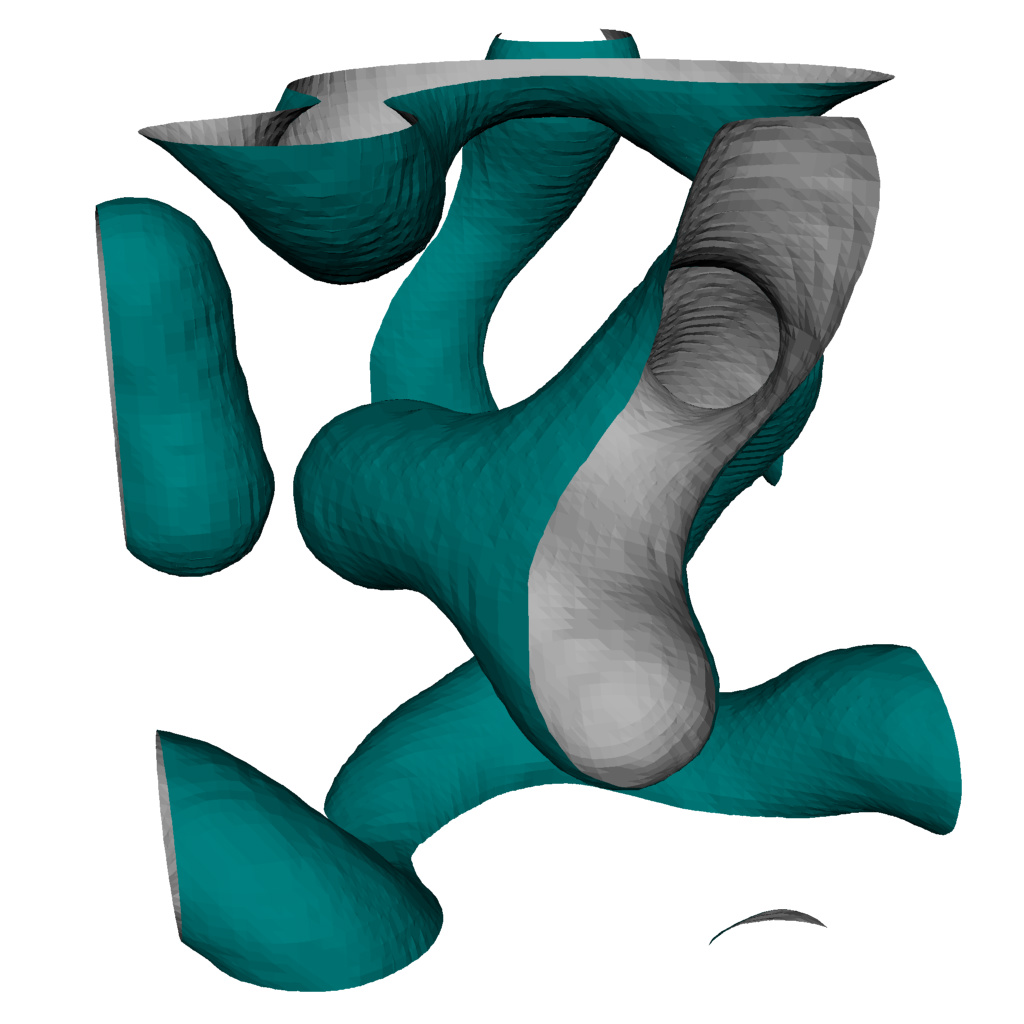}
		\label{fig:3DPatternsB1d}
	}
	\hfill
	\subfigure[\colorbox{gray!50}{\textcolor{sphere0}{UC spheres $\blacksquare$}}]{
		\includegraphics[width=.17\textwidth]{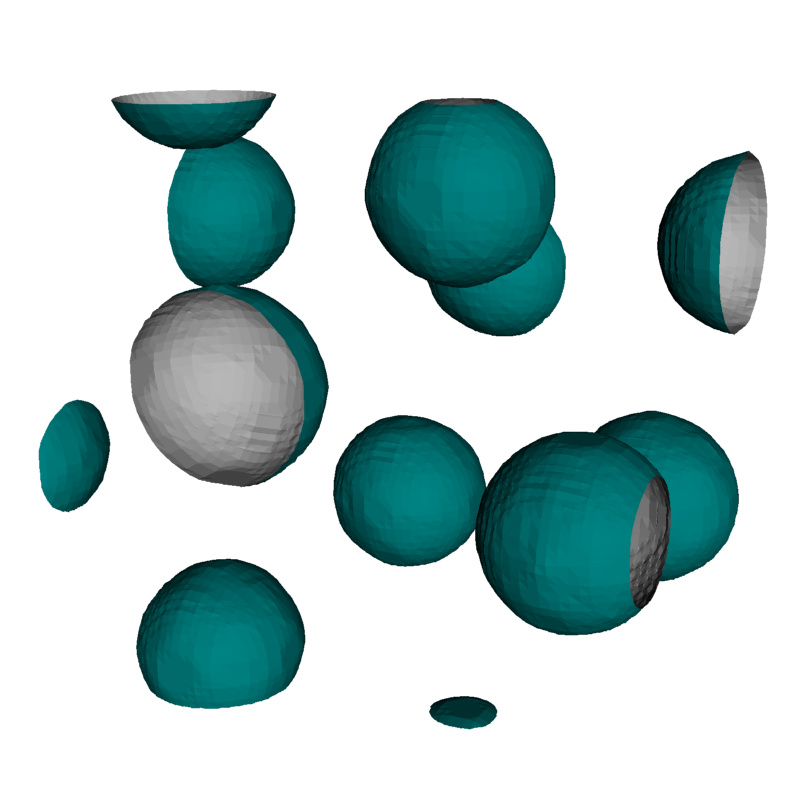}
		\label{fig:3DPatternsB0d}
	}
	\caption{Five of the seven possible Turing-like patterns in a three-torus, with the trivial cases of all DCs or all UCs not shown. Patterns include $ 0 $-dimensional, sphere-like structures ((a) and (e)), $ 1 $-dimensional, pipe-like structures ((b) and (d)), and $ 2 $-dimensional, area-spanning structures (c).}
	\label{fig:3DPatterns}
\end{figure}

The three parameters~$\rho$, $R_1$, and $R_2$ span a~$ {121 \times 40\times 40} $ grid where each including triple gives rise to a Turing-like pattern.
Based on the works~\cite{leppanen2002new,bansagi2011tomography,skrodzki2017turing}, we expect to find seven different types of structures that can form in the three-dimensional domain.
Two of these are the trivial patterns of only differentiated or only undifferentiated cells.
The remaining five structures are displayed in Figure~\ref{fig:3DPatterns}. 
We can find~$ 0 $-dimensional structures (more or less spheres, see Figures~\ref{fig:3DPatternsW0d} and \ref{fig:3DPatternsB0d}), $ 1 $-dimensional structures (pipe-like, Figures~\ref{fig:3DPatternsW1d} and \ref{fig:3DPatternsB1d}), and $ 2 $-dimensional structures (clear flat, planar areas visible, Figure~\ref{fig:3DPatternsWB2d}). 
Observe, that except for the $ 2 $-dimensional case, we obtain two possible scenarios for $0$-dimensional and $1$-dimensional structures, which come from the distinction into interior and exterior cell types.
That is, for instance the spheres could be made from DCs while being surrounded by UCs (Figure~\ref{fig:3DPatternsW0d}) or they could contain the UCs while being surrounded by DCs (Figure~\ref{fig:3DPatternsB0d}).
An analogous distinction holds for the $1$-dimensional patterns.
The experimental part of the project now consists of partitioning the discretized parameter space into these seven cases, i.e.\@ identifying for each parameter triple~$ {(\rho,R_1,R_2)} $ which type of Turing-like pattern arises.

\subsection{The experiment}
\label{sec:TheExperiment}

To label all triples in the discretized parameter space, we set up a citizen science project.
It was run during the ``Long Night of Sciences''\footnote{German Website: https://www.langenachtderwissenschaften.de/} in 2019.
This is an annual public event in several cities in Germany taking place since 2001. 
Scientific institutions open their doors and present (hands-on) research exhibits to the public audience for young and old alike.
At our booth, visitors were first introduced to three-dimensional Turing-like pattern, via the results of~\cite{skrodzki2017turing}.
Furthermore, they were acquainted with the different non-trivial three-dimensional patterns via images corresponding to those in Figure~\ref{fig:3DPatterns}.
Being thus prepared, the visitors could contribute to the exploration of the parameter space.

\begin{figure}
	\centering
	\includegraphics[width=.40\textwidth]{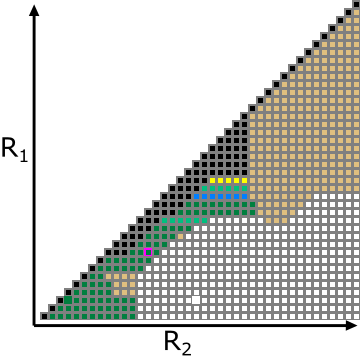}
	\hspace*{1cm}
	\includegraphics[width=.40\textwidth]{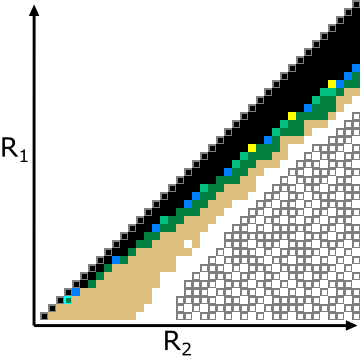}
	\caption{Each choice of~$\rho$ reduces the parameter-space to two dimensions, where the $x$-axis represents~$R_2$ and the $y$-axis represents~$R_1$ with~$(1,1)$ in the bottom left corner. Squares with a gray border indicate not yet classified triples, those without a border are classified, the pink border indicates the currently selected triple. Colors indicate either the classified color of the triple or the type of the closest triple in the three-dimensional parameter space, following the color coding introduced in Figure~\ref{fig:3DPatterns}. Note that all pairs equal and above the black diagonal represent scenarios with~$ {R_1\geq R_2} $ which result in all cells being differentiated. Both images display the layer $ {\rho = \sigma(24)} $, where the left represents estimates during the progress and the right represents the final stage of our experiments for this slice of the parameter space. Note that the bottom right corner with majorly unvisited cells will generate ``UC all'' structures.}
	\label{fig:ParameterSpace}
\end{figure}

In our software on display during the event, the visitors would first choose a value for the initial differentiation probability~$\rho$ on the scale given by~$ {\mathcal{I}=\{0,\ldots,121\}} $.
Making this choice reduces the three-dimensional to a two-dimensional parameter space.
This lower-dimensional parameter space was then shown to the visitors, see Figure~\ref{fig:ParameterSpace}.
From this display, a pair~$(R_1,R_2)$ is selected and the corresponding Turing-like pattern is generated iteratively following Equation~(\ref{equ:TuringLikeIteration}).
After the automaton has either converged (i.e.\@ no changes occurred after a time step) or a maximal number of iterations or seconds has passed, the current state is rendered using the \emph{Marching Cubes Algorithm} as described by~\cite{skrodzki2017turing}.
The visitors were then asked to either rate the resulting pattern according to the images in Figure~\ref{fig:3DPatterns} or (if the pattern clearly had not converged yet) to continue the generation for several more iterations.

While the citizen science approach harnesses the power of the masses in evaluating as many triples as possible, it also bears problems.
On the one hand, the visitors to the exhibition were not continuously monitored, thus they could deliberately mis-classify triples of the parameter space.
On the other hand, as some of the patterns take many iterations to converge, visitors might unintentionally mis-classify a not yet fully converged pattern.
After the initial classification during the public science event, we thus had to perform some cleaning operations to re-classify outliers in the parameter space.
Furthermore, we classified as many remaining triples as necessary by hand to identify the separating surfaces between connected volumes in the three-dimensional parameter space, corresponding to the five different patterns as shown in Figure~\ref{fig:3DPatterns}.
The relevant part of the three-dimensional parameter space still contains~$ {121\cdot40\cdot39 / 2=94,380} $ triples.
Thus, a full---visually guided---partitioning is only possible with an efficient implementation.

\subsection{Implementation details}

\begin{figure}
	\centering
	\subfigure[]{
		\includegraphics[width=.2\textwidth]{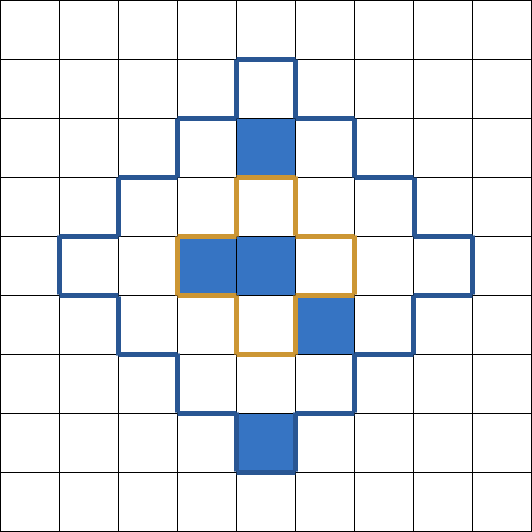}
		\label{fig:evalAreas}
	}
	\hfill
	\subfigure[]{
		\includegraphics[width=.2\textwidth]{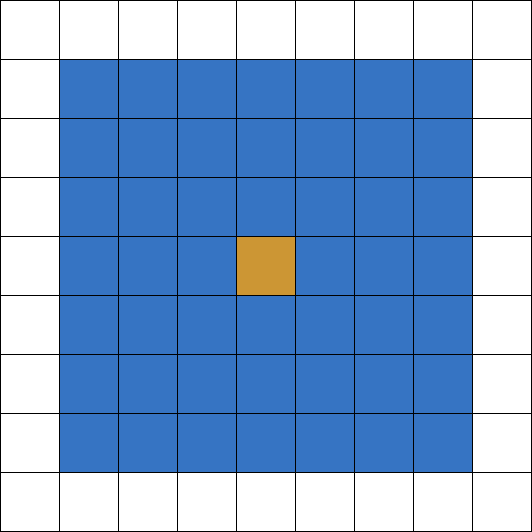}
		\label{fig:evalBox}
	}
	\hfill
	\subfigure[]{
		\includegraphics[width=.2\textwidth]{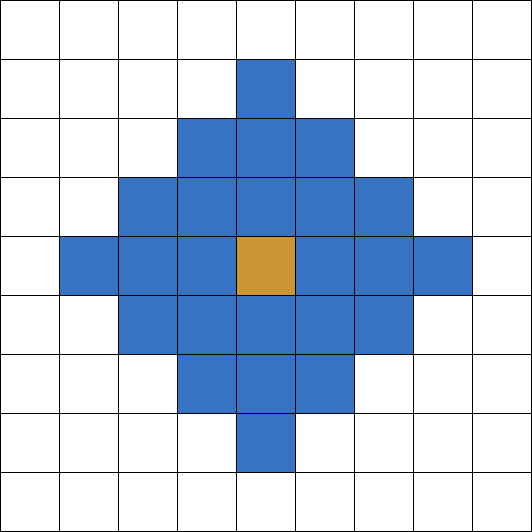}
		\label{fig:evalSphere}
	}
	\hfill
	\subfigure[]{
		\includegraphics[width=.2\textwidth]{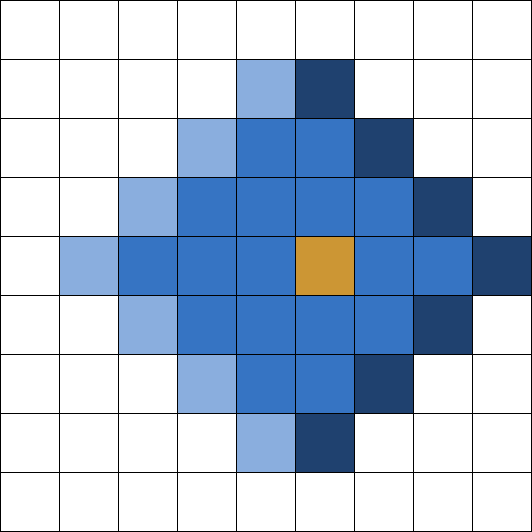}
		\label{fig:evalDiffKernel}
	}
	\caption{Two-dimensional illustrations of our implementation of the kernel evaluation. (a) Inner kernel of radius~$R_1$ (yellow) and outer kernel (annulus in the two-dimensional setup) with radii~$R_1$ and~$R_2$ (blue) including several DCs. (b) Box of size~$ {R_2\times R_2} $ as naive evaluation of the kernel. (c) Exact evaluation of the kernel by computing involved cell indices beforehand. (d) Evaluation of the difference when shifting the kernel keeps most information constant and only causes evaluation on the edge of the kernel.}
	\label{fig:evalulationAndKernels}
\end{figure}

We implemented the iterative procedure given by Equation~(\ref{equ:TuringLikeIteration}) within the geometry visualization framework \emph{JavaView} by~\cite{polthier2020javaview}.
The framework is based on \emph{Java} and provides several comfort functions and interaction tools~\cite[cf.][Sec.~2]{polthier2002publication}.
When starting the procedure for a given triple~$ {(\rho,R_1,R_2)} $, a first initial choice of differentiated cells is performed, uniformly random according to~$ \rho $.
Then, iteratively, Equation~(\ref{equ:TuringLikeIteration}) is evaluated for each cell. 
Therefore, the spherical kernel with radius~$ R_1 $ and the larger ball-cavity kernel with outer radius~$ R_2 $ and inner radius~$ R_1 $ are evaluated for each cell see Figure~\ref{fig:evalAreas} for a two-dimensional illustration.
This evaluation consists of counting the DCs within radius~$ R_1 $ of the center cell (shown in yellow) and weighting this number against the number of DCs that lie within radius~$ R_2 $ of the center cell, but farther away than~$ R_1 $.
The center cell will then be a DC in the next time step as prescribed by Equation~(\ref{equ:TuringLikeIteration}).

The calculation of one time-step of the cubical grid is time consuming. 
As reasoned above, our cubical size choice was a justifiable trade-off between obtainable structure size and calculation time. 
A naive implementation would consider a box of size~$ {R_2\times R_2\times R_2} $ around the center cell and then compute for each cell whether it should be considered in either kernel, see Figure~\ref{fig:evalBox} for a two-dimensional illustration.
By computing the kernels beforehand, it is possible to drop~$ {R_2^3 - \frac{4\pi}{3}R_2^3} $ operations, causing for a significant speedup, see Figure~\ref{fig:evalSphere} for a two-dimensional illustration.
However, we can further enhance the computation by considering a spherical difference area.
That is, when we switch from one central cell to the next and query its neighboring cells, we only consider those that got changed w.r.t.\@ the kernel shift from one cell to the other, see Figure~\ref{fig:evalDiffKernel} for a two-dimensional illustration.
This reduces the number of cells to be evaluated to~$ {4\pi (R_2^2+R_1^2)} $.

A last optimization was achieved via parallelization.
We split the cubical domain in chunks up to the number of available threads and process them in parallel. 
The machine used in the citizen science setup described above used 24 threads for parallel processing of the iterations.
We also employ parallelization for the Marching Cubes isosurface construction for the visualization of the patterns.
However, even with these improvements, timely convergence of the patterns is not guaranteed. 
To ensure the user having an interactive experience, we further integrated an iteration and a time limit, as indicated above, such that the computation will automatically stop after convergence or after one of the given limits is reached. 
In the experimental setup, computation was halted after~$ 20 $ iterations or seconds respectively, if no convergence was reached yet.
Following the work of~\cite{schwehm2016fast}, the generation of the patterns as well as their visualization could be sped up further by utilizing GPU computation.

\subsection{The results}

\begin{figure}
	\centering
	\subfigure{
		\includegraphics[width=.47\textwidth]{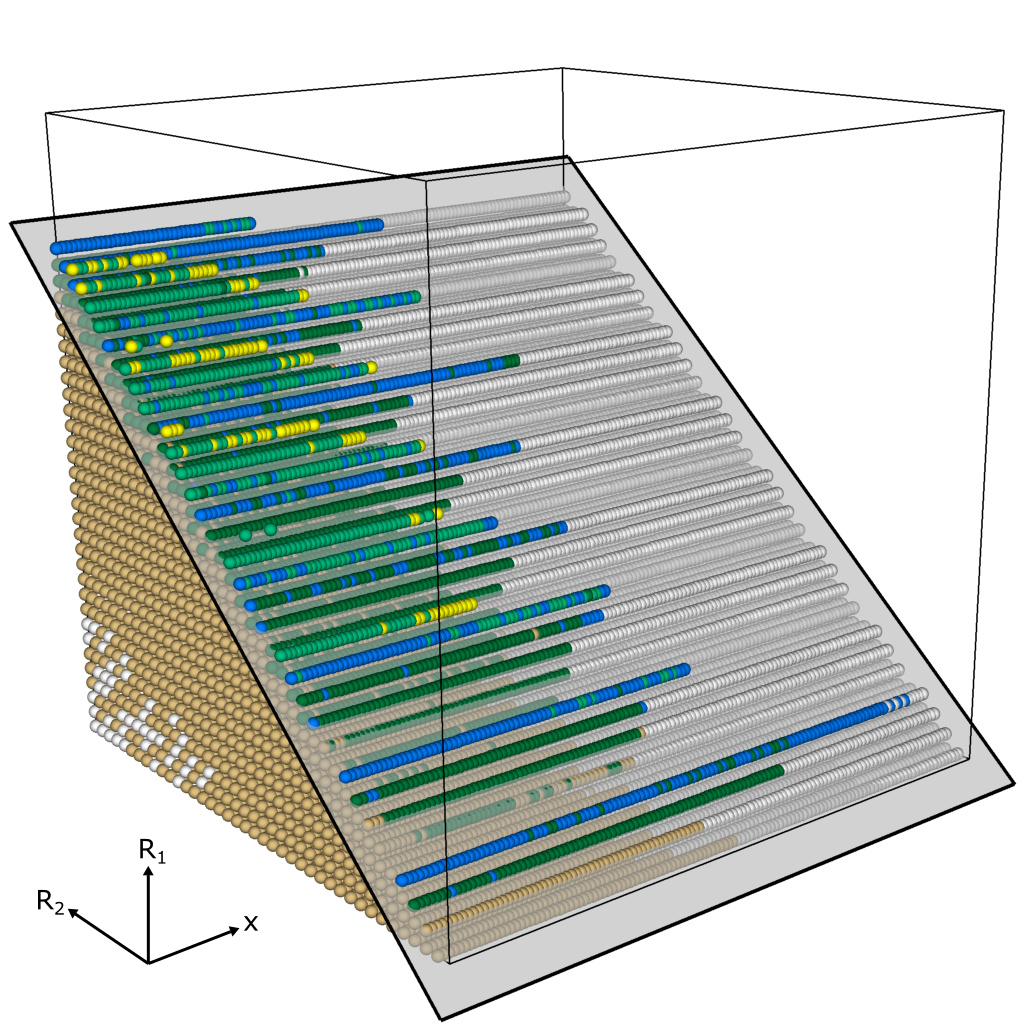}
		\label{fig:3dResults}
	}
	\hfill
	\begin{tikzpicture}
	\begin{axis}[
		symbolic x coords={DC all,DC spheres,DC pipes,areas,UC pipes,UC spheres,UC all},
		x tick label style={rotate=45, anchor=east, align=left, font=\tiny}, %
		ybar, 
		bar shift=0pt, 
		ymode=log, 
		height=0.45\textwidth, width=0.54\textwidth, 
		nodes near coords, 
		every node near coord/.append style={font=\tiny},
		nodes near coords align={vertical},
		point meta=rawy, 
	]
	\addplot[black,fill=black] coordinates {(DC all,116432)};
	\addplot[black,fill=sphere1] coordinates {(DC spheres,10348)};
	\addplot[black,fill=tube1] coordinates {(DC pipes,2218)};
	\addplot[black,fill=surface] coordinates {(areas,607)};
	\addplot[black,fill=tube0] coordinates {(UC pipes,408)};
	\addplot[black,fill=sphere0] coordinates {(UC spheres,98)};
	\addplot[black,fill=white] coordinates {(UC all,63489)};
	\end{axis}
	\end{tikzpicture}    
	\caption{Left: Plot of 3D space consisting of activator radius~$ R_1 $, inhibitor radius~$ R_2 $, and activation ($ x $ as argument for $ {\rho = \sigma(x)} $). The parameter triples are colored by found structures. The shown hyperplane with increase~$ \sqrt[3]{2}^{-1} $ indicates for what elements of the parameter space the volumes of inhibitor and activator kernels are equal. Note that the color for all-DC triples is black, and we disabled their rendering for enhanced visibility of the other structures, cf.~Figure~\ref{fig:surface1_2}. Note that the $x$-axis in the left picture is not a linear depiction of the activation probability, but the argument for~$\sigma(x)$, see Equation~(\ref{equ:sigmoid}). Right: Histogram over the found structures. Note that the $y$-axis in the histogram follows a logarithmic scale. Both figures follow the color coding according to Figure~\ref{fig:3DPatterns}.}
	\label{fig:3dresults_hist}
\end{figure}

In this section, we are going to evaluate the experimentally found structures and how they are distributed. 
According to the histogram in Figure~\ref{fig:3dresults_hist}, the dominating structures are ``DC all'' followed by ``UC all'', which enclose all other structures in the parameter space. 
This encapsulating behavior can be seen in the two-dimensional layers of the parameter space shown in Figure~\ref{fig:ParameterSpace} and in the left image of Figure~\ref{fig:3dresults_hist}, showing the entire discretized parameter space. 
Note that in Figure~\ref{fig:3dresults_hist} and in the upcoming illustrations, we decided to neglect the ``DC all'' triples for visual reasons: Rendering these would hide the distribution of the other structures within the parameter space.
Going back to the histogram in Figure~\ref{fig:3dresults_hist}, the remaining non-trivial structures follow with decreasing amount, with ``UC pipes'' and ''UC spheres'' having the rarest occurrence.

One first trivial observation about the distribution of structures can be drawn from the behavior of the activator and inhibitor radii: If $ {R_1 \geq R_2} $, all cells are differentiated (given that the random process guided by~$ {\rho>0} $ creates at least one active cell), because the inhibitor range does not have any impact.
Therefore, all triples equal to and above the diagonal in Figure~\ref{fig:ParameterSpace} can be classified as ``DC all''.

Considering the three-dimensional parameter space and the distribution of structures in it, as shown in Figure~\ref{fig:3dresults_hist}, suggests another separating plane. 
The majority of ``areas'', ``UC pipes'', and ``UC/DC sphere'' structures resides close to this plane.
When considering the three-dimensional kernel as defined in Section~\ref{sec:SteppingUpOneDimension} and the special case of inhibitor and activator volume to be equal, we obtain:
\begin{equation*}
	\text{vol(activator)}=\text{vol(inhibitor)}\quad \Rightarrow\quad\frac{4}{3}\pi R_1^3 = \frac{4}{3}\pi R_2^3 - \frac{4}{3}\pi R_1^3\quad \Rightarrow \quad R_1 = \frac{R_2}{\sqrt[3]{2}}.
\end{equation*} 
Therefore, in each probability layer (i.e.\@ for a fixed value~$\rho$) we can draw a separating line with increase $ \sqrt[3]{2}^{-1} $. 
Respectively, we can separate the entire three-dimensional parameter space by a separating hyperplane, see Figure~\ref{fig:3dResults}.
Above this separation, the triples are more likely to produce ``DC all'' structures and below it, they tend to form ``UC all'' structures. 
However, as we are dealing with a discretization---i.e.\@ the volume of kernels is not~$ {\frac{4}{3}\pi R_i^3} $,~$ {i=1,2} $ but the number of discrete cells within---the obtained structures in our experiments deviate from this theoretical assertion.
All such structures (``areas'', ``UC pipes'', and ``spheres'') occur in the vicinity of this linear separation, see Figure~\ref{fig:3dresults_hist}).

\begin{figure}
	\centering
	\subfigure{
		\includegraphics[width=.47\textwidth]{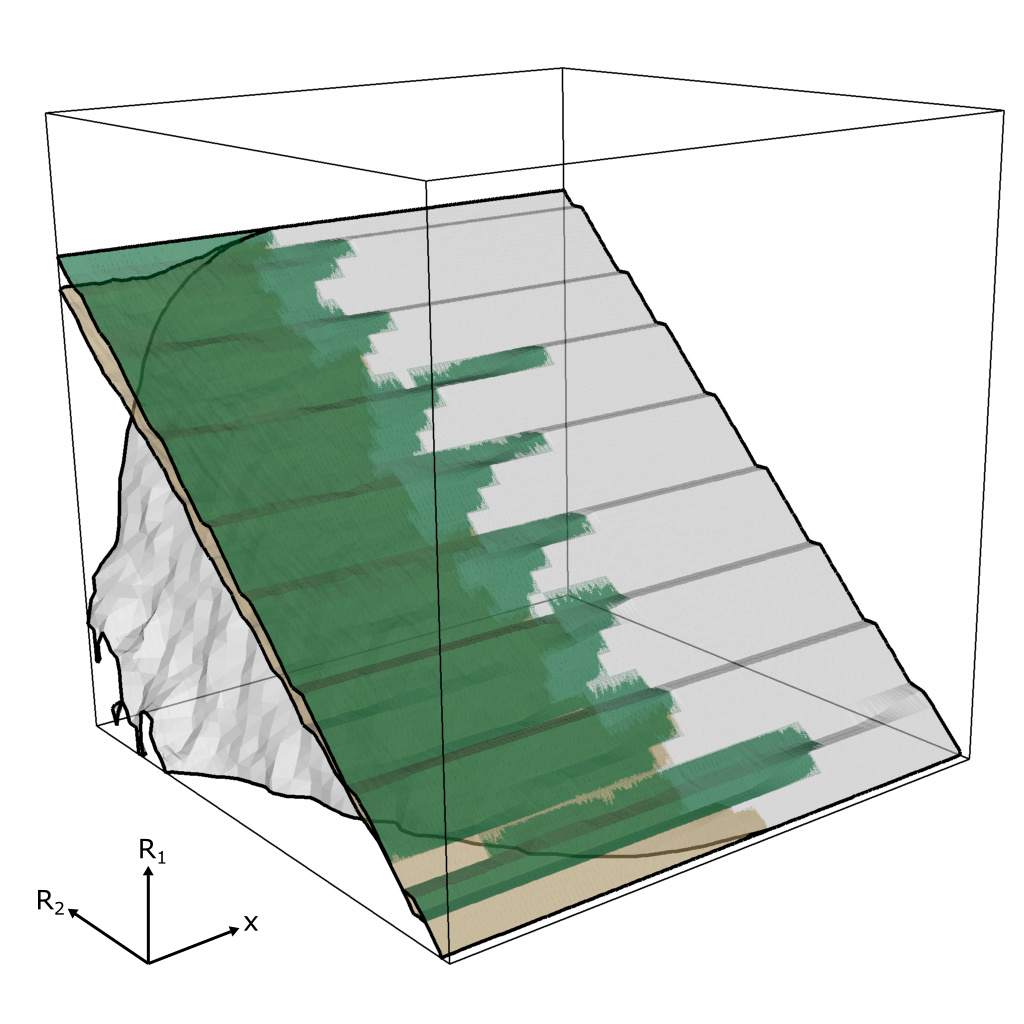}
		\label{fig:surface1}
	}
	\hfill
	\begin{minipage}[b]{0.47\textwidth}
		\includegraphics[width=.47\textwidth]{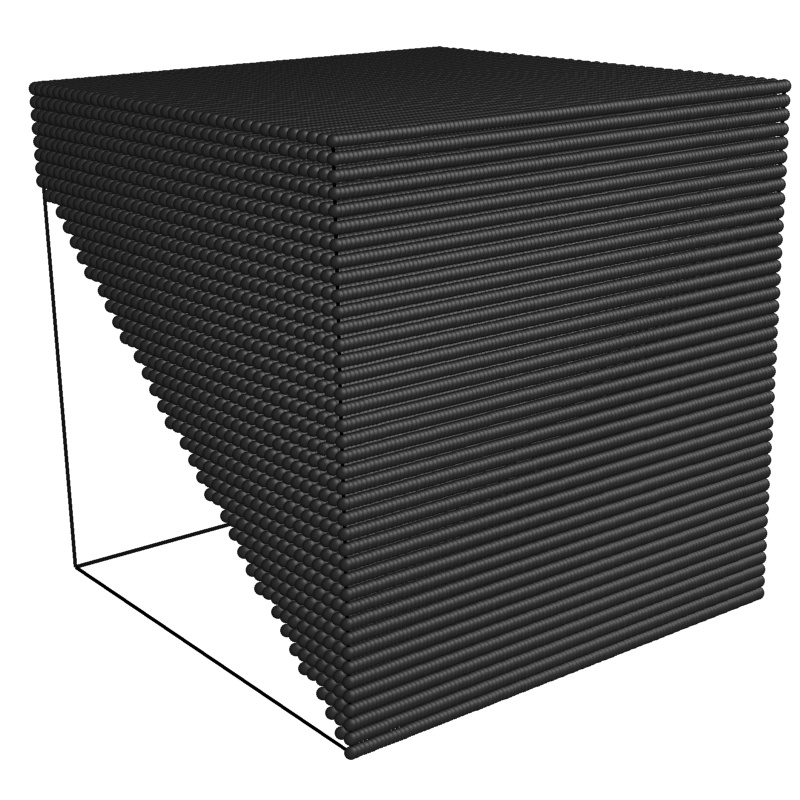}
		\hfill
		\includegraphics[width=.47\textwidth]{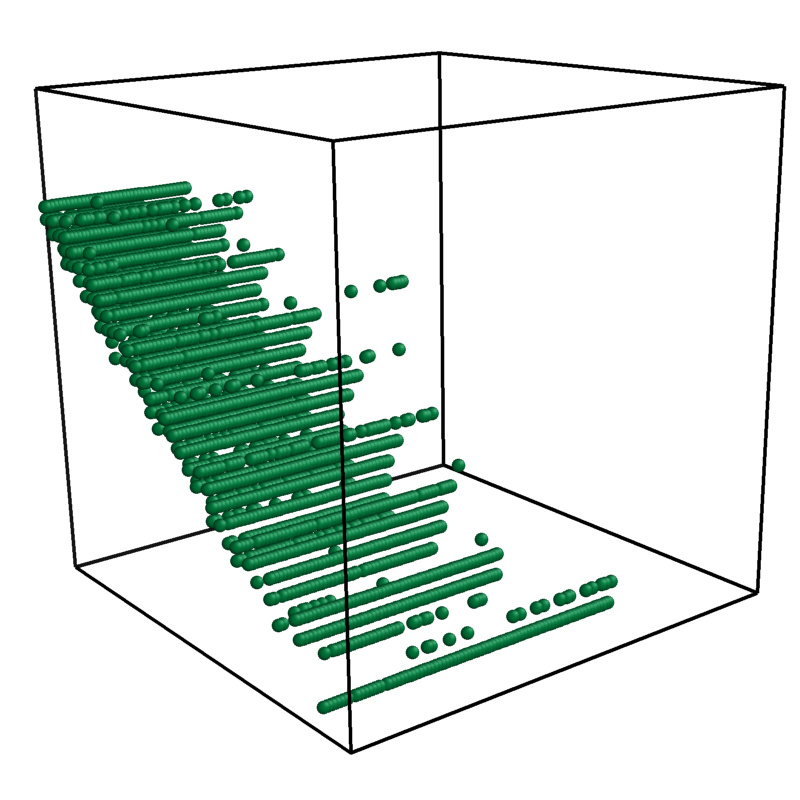}
		\includegraphics[width=.47\textwidth]{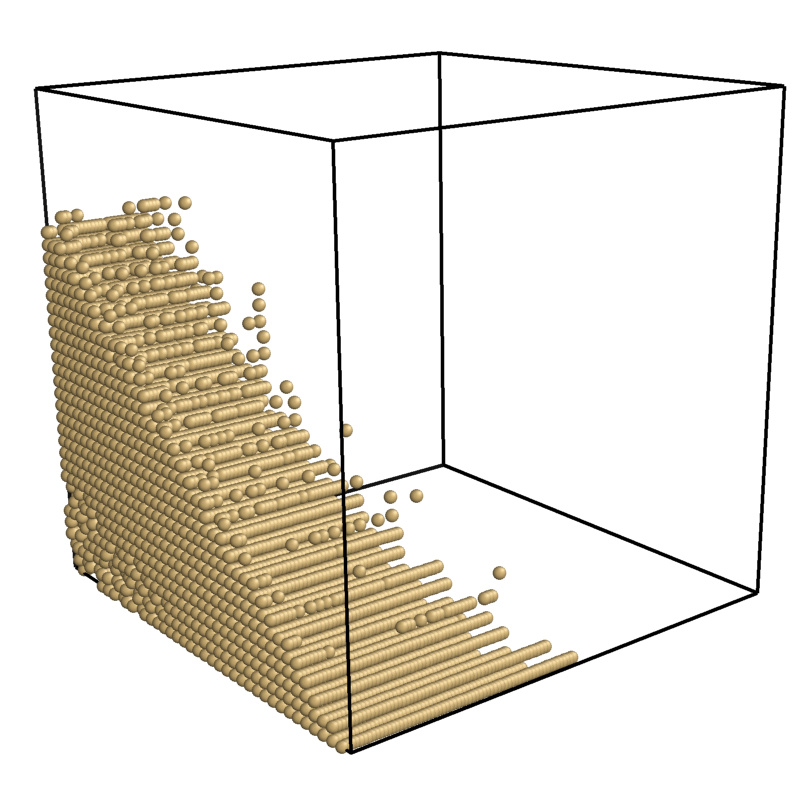}
		\hfill
		\includegraphics[width=.47\textwidth]{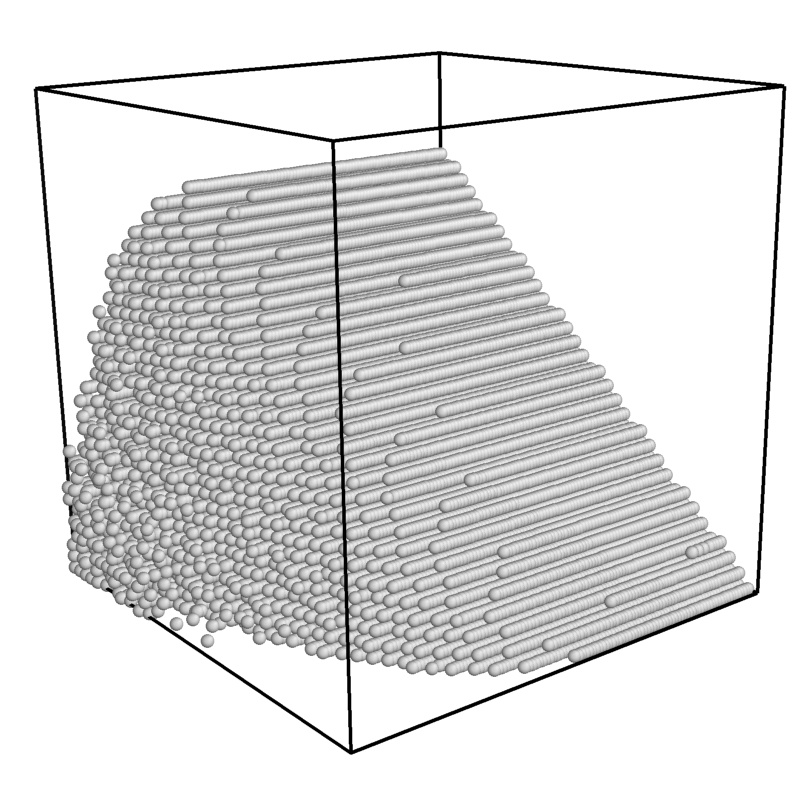}
	\end{minipage}
	\caption{Left: Separating isosurfaces of volumetric distributed triples ``UC all'' (behind white), ``DC spheres'' (between white and beige), ``DC pipes'' (between beige and green), and ``DC all'' (remaining volume on top). Right: The respective triple groups from the parameter set, giving rise to the separating isosurfaces, i.e.\@ ``DC all''~$\blacksquare$, ``DC pipes''~\textcolor{tube1}{$\blacksquare$}, ``DC spheres''~\textcolor{sphere1}{$\blacksquare$}, and ``UC all'' in white.}
	\label{fig:surface1_2}
\end{figure}

Within our discretization of the parameter space, we find that those triples form volumetric groups that give rise to the following four structures: ``UC all'', ``DC all'', ``DC pipes'', and ``DC spheres''.
On the left in Figure~\ref{fig:surface1_2}, we draw separation areas (as isosurfaces) in white, beige, and dark green that separate the respective triple groups.
The groups are shown on the right in Figure~\ref{fig:surface1_2}.
That is, most of the cube is filled with ``DC all'' triples, filling it from the top.
From the dark green up to the beige isosurface, we find a slender wedge consisting of ``DC pipes''.
The volume between the beige isosurface and the highly non-planar white surface is filled with ``DC spheres'' while the rest of the parameter space consists of ``UC all''.
The other structures (``areas'', ``UC pipes'', and ``UC spheres'') do not obtain volumetric extent in our discretization and with the chosen parameter bounds. 
They all lie close to the linear separation plane shown in Figure~\ref{fig:3dresults_hist}.
These observations motivate two conjectures.
\begin{conj}
	The wedge, formed by the ``DC pipes'' triples, given as volume between the beige and dark green isosurface, will grow larger and wider with increasing values for~$R_1$ and~$R_2$.
\end{conj}
\begin{conj}
	The structures: ``areas'', ``UC pipes'', and ``UC spheres'' will gain volumetric extent when further increasing the values for~$R_1$ and $R_2$.
\end{conj}

\begin{figure}
	\centering
	\includegraphics[width=.49\textwidth]{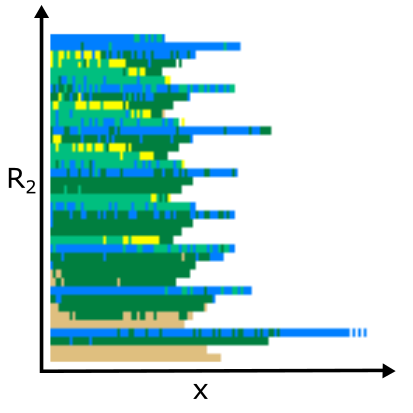}
	\hfill
	\begin{minipage}[b]{0.49\textwidth}
		\includegraphics[width=1.\textwidth]{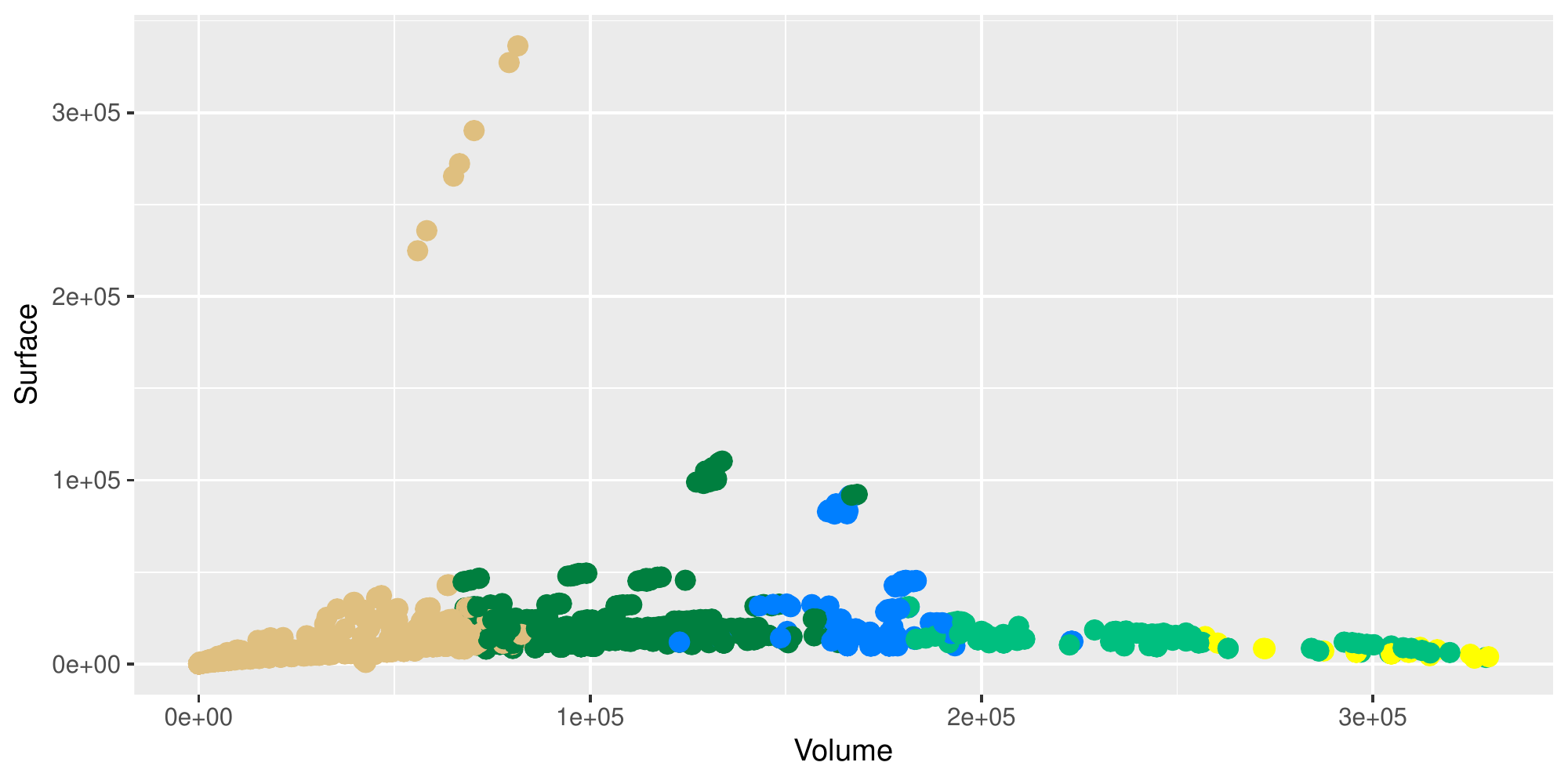}
		\includegraphics[width=1.\textwidth]{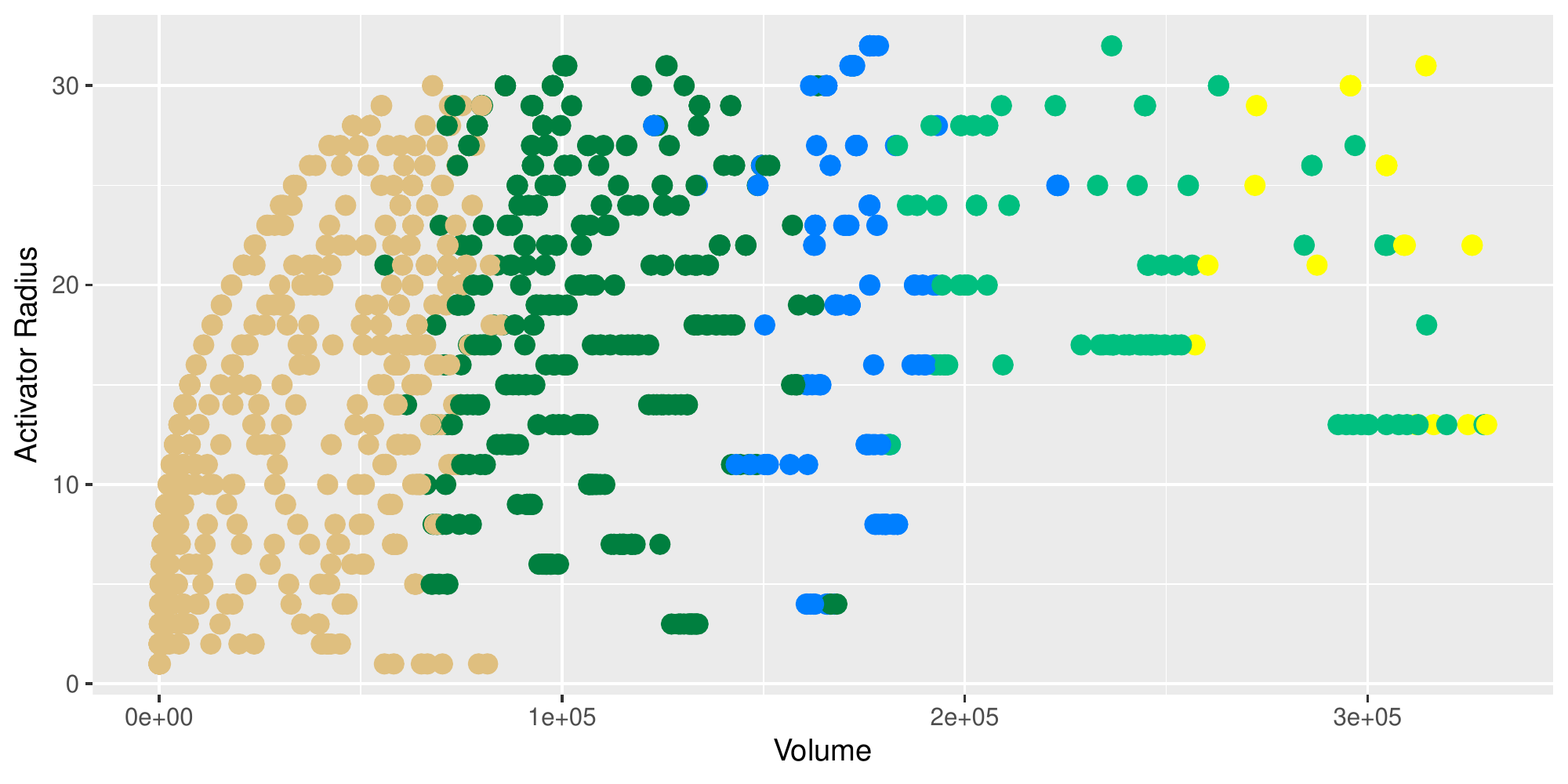}
	\end{minipage}
	\caption{Left: A top view of the left image in Figure~\ref{fig:3dresults_hist}, i.e.\@ a projection along the $R_1$-axis. Each square represents the coordinate triple with corresponding $x$ and $R_2$ value and the color of the largest $R_1$ triple that does not represent ``DC all''. Right: Two scatterplots indicating the relationship between volume and surface area (top) as well as volume and activator radius (bottom) of the sub-structures.}
	\label{fig:resTopViewAndScatterplots}
\end{figure}

Figure~\ref{fig:resTopViewAndScatterplots} illustrates the top view, showing the occurrence and distributions of those structures lying close to the separating plane shown in Figure~\ref{fig:3dresults_hist}.
Note how the blue ``area'' structures exhibit an almost linear behavior when increasing~$\rho$ while keeping~$ R_1 $ and~$ R_2 $ fixed. 
These blue spikes exhibit almost equidistant starting points. 

The ``UC spheres'' shown in yellow also exhibit a curve-like behavior, not as straight as the blue ``area'' structures, though.
They start below the blue lines and gradually step down towards the next blue line with growing values of~$\rho$.
The ``UC pipes'' shown in light green, however, behave different as they clearly extend over more than just curves, but rather span the two-dimensional space between blue and yellow lines. 
Note that with low~$ R_1, R_2 $ values (towards the bottom in the left Figure~\ref{fig:resTopViewAndScatterplots}) we do not find ``UC pipes'' or ``UC spheres'', which might however be caused by the chosen resolution of our discretization.

Consequently, we were able to find all structures displayed in Figure~\ref{fig:3DPatterns}.
Thereby, we positively answer the corresponding conjecture, posed in~\cite{skrodzki2017turing}.
Our experiments lead us to the following conjecture.
\begin{conj}
	A Turing-like pattern of dimension~$ d $ exhibits sub-structures of all dimensions from~$ 0 $ to~$ {d-1} $.
	Furthermore, each structure of dimension~$ {0, \ldots, d-2} $ occurs twice as ``UC'' and ``DC'' version, while sub-structures of dimension~$ {d-1} $ appear only once.
\end{conj}

\begin{figure}
	\begin{tikzpicture}
		\begin{axis}[
			width = \textwidth,
			height = 0.5\textwidth,
			ybar stacked,
			bar width=10pt,
			ymin=0, 
			ymax=85, 
			xmin=1,
			xmax=20,
			restrict y to domain*=0:95, 
			xtick=data,
			enlarge x limits=0.03,
			symbolic x coords={1, 2, 3, 4, 5, 6, 7, 8, 9, 10, 11, 12, 13, 14, 15, 16, 17, 18, 19, 20},
			after end axis/.code={ 
				\draw [ultra thick, white, decoration={snake, amplitude=1pt}, decorate] (rel axis cs:0,1.0) -- (rel axis cs:1.0,1.05);
			}, 
			axis lines*=left, 
			clip=false, 
			xlabel = {Percentage of active cells}, 
			xlabel style={yshift=-1cm},
			x tick label style={rotate=90}, 
						xticklabels = {
				{$(0,5]$},
				{$(5,10]$}, 
				{$(10,15]$}, 
				{$(15,20]$},
				{$(20,25]$},
				{$(25,30]$},
				{$(30,35]$},
				{$(35,40]$},
				{$(40,45]$},
				{$(45,50]$},
				{$(50,55]$},
				{$(55,60]$},
				{$(60,65]$},
				{$(65,70]$},
				{$(70,75]$},
				{$(75,80]$},
				{$(80,85]$},
				{$(85,90]$},
				{$(90,95]$},
				{$(95,100)$}},
		]
		\addplot[fill=sphere1,draw=black,nodes near some coords={0/{213}/above}] coordinates {
			(1,213) (2,71) (3,72) (4,88) (5,51) (6,0) (7,0) (8,0) (9,0) (10,0) (11,0) (12,0) (13,0) (14,0) (15,0) (16,0) (17,0) (18,0) (19,0) (20,0)};
		\addplot[fill=tube1,draw=black,nodes near some coords={3/{169+7}/above left,4/{51+68}/above, 5/{141}/above right}] coordinates {
			(1,0) (2,0) (3,0) (4,7) (5,44) (6,141) (7,67) (8,63) (9,28) (10,11) (11,0) (12,0) (13,0) (14,0) (15,0) (16,0) (17,0) (18,0) (19,0) (20,0)};
		\addplot[fill=surface,draw=black] coordinates {
			(1,0) (2,0) (3,0) (4,0) (5,0) (6,0) (7,0) (8,3) (9,9) (10,49) (11,73) (12,6) (13,1) (14,6) (15,0) (16,0) (17,0) (18,0) (19,0) (20,0)};
		\addplot[fill=tube0,draw=black] coordinates {
			(1,0) (2,0) (3,0) (4,0) (5,0) (6,0) (7,0) (8,0) (9,1) (10,0) (11,8) (12,24) (13,6) (14,12) (15,26) (16,3) (17,3) (18,13) (19,4) (20,1)};
		\addplot[fill=sphere0,draw=black] coordinates {
			(1,0) (2,0) (3,0) (4,0) (5,0) (6,0) (7,0) (8,0) (9,0) (10,0) (11,0) (12,0) (13,0) (14,0) (15,1) (16,4) (17,1) (18,11) (19,8) (20,3)};
		\end{axis}
	\end{tikzpicture}
	\caption{Histogram over 2,214 data points from the parameter space, selected from the border of two sub-structure areas. The bins on the $x$-axis represent the percentage of activated cells in the final state (excluding ``UC all'' and ``DC all'' at the far ends). The stacked bars are colored following the color scheme of Figure~\ref{fig:3DPatterns}. Bars are printed up to 85 elements and cut from there, where the actual values are given on top.}
	\label{fig:StructureHistogram}
\end{figure}
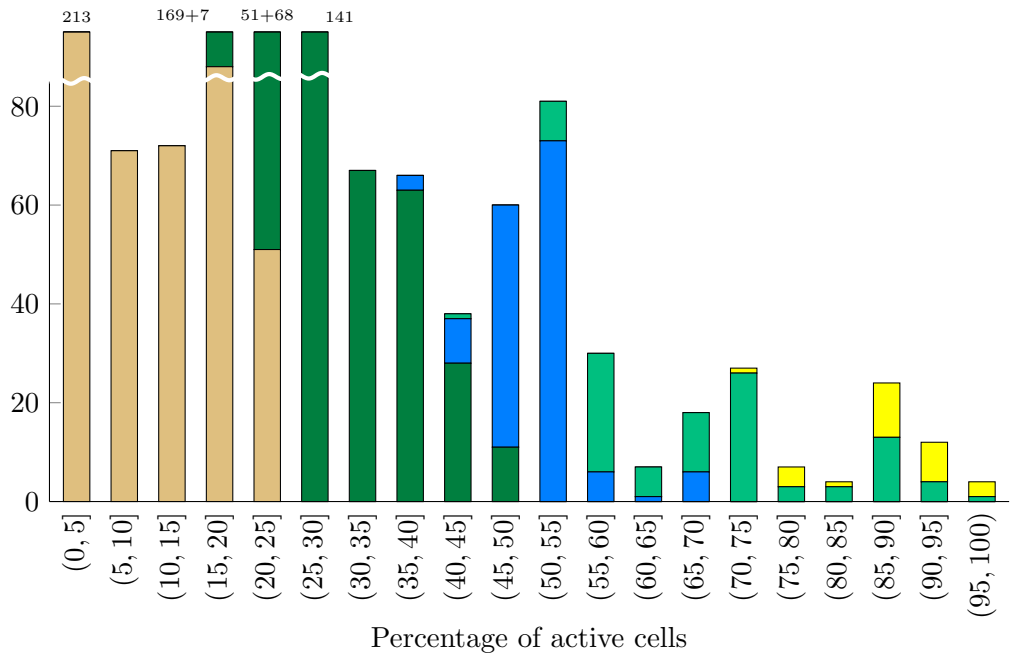

\begin{figure}
	\includegraphics[width=1.\textwidth]{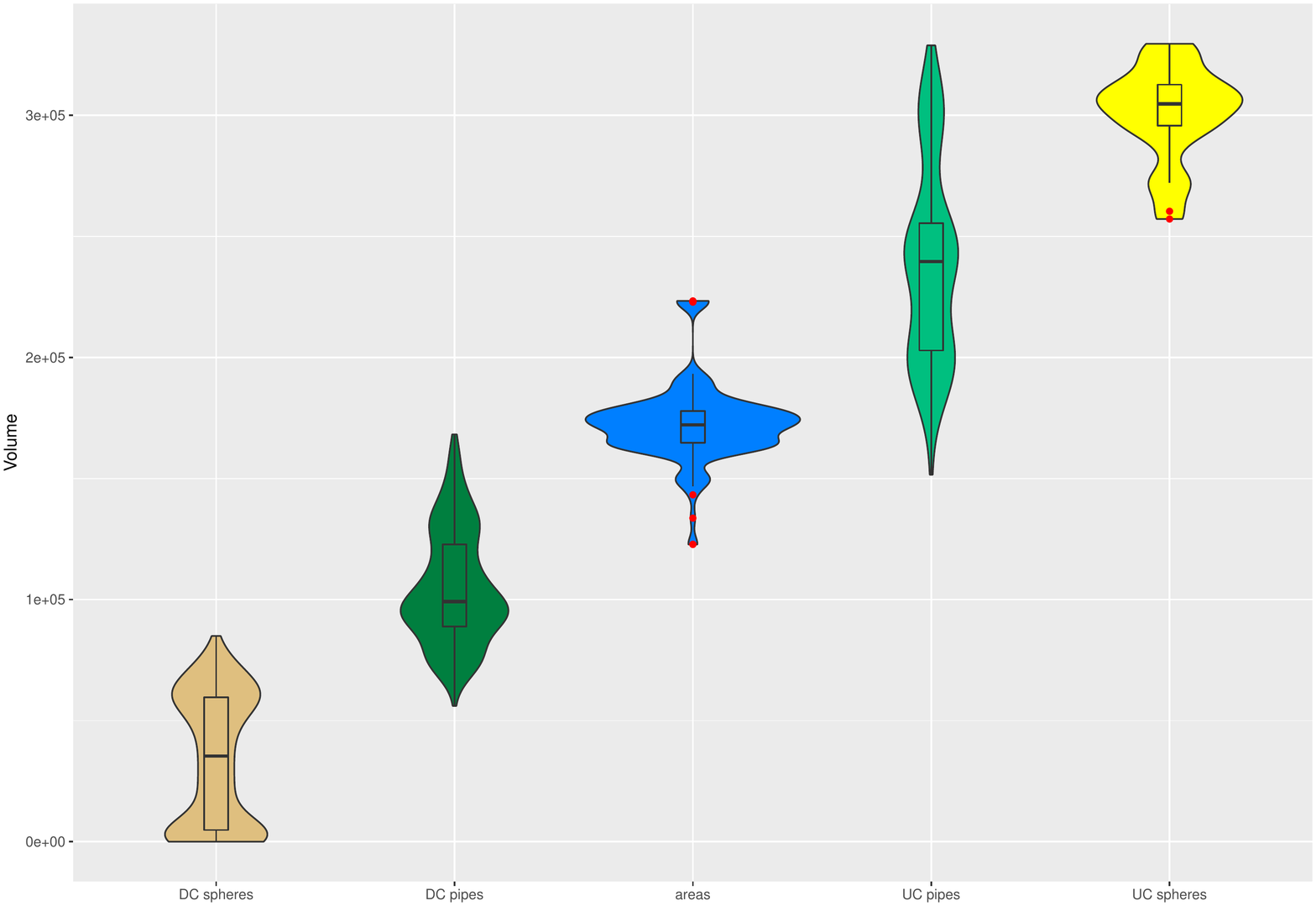}
	\caption{Violin plot over the different types of sub-structures, indicating the volume that is taken by each sub-structure. Note the box-plot overlay indicating the mean as well as the first and third quartile. Outliers are shown in red, most notably for the ``areas''. Furthermore, ``UC pipes'' and ``UC spheres'' do not show a clear separation by their volume.}
	\label{fig:ViolinPlotVolume}
\end{figure}

Finally, we are going to use the obtained data in order to better classify the different sub-structures and to work towards a thorough mathematical description of these sub-structures.
The scatterplots shown in Figure~\ref{fig:resTopViewAndScatterplots} suggest that the sub-plots are characterized to a large extend via their volume.
Indeed, a corresponding histogram (Figure~\ref{fig:StructureHistogram}) and a violin plot (Figure~\ref{fig:ViolinPlotVolume}) confirm this first suspicion.

Based on these preliminary analyses, we aim at building a predictor for the sub-structure type, based on the volume found.
We use a total of~$2,214$ hand-classified sub-structures to train the predictor.
Note that this data is not uniformly selected, but rather chosen along the border of larger patches as shown in Figure~\ref{fig:ParameterSpace}.
For cross-correlation, we separate the data set into five folds, with the distribution of sub-structures as shown in Table~\ref{tab:FoldDistribution}.
To train our predictor, we perform the following operation on all folds.
For two neighboring sub-structure types, we use the training data (all data points not in the fold) to find a certain volume-threshold to optimally distinguish these.
Thus, we obtain four volume-thresholds for each fold, indicating five intervals that are associated with the respective sub-structures.
Then, we use these thresholds to classify the testing data from the fold.
Table~\ref{tab:FoldDistribution} indicates the classification error made within each fold.
Overall, the prediction of the sub-structure by the volume admits an average classification error of~$8.39\%$.
While this indicates the volume to be a very strong predictor, the results differ drastically over the different sub-structures.
While ``DC spheres'' exhibit an extremely low classification error of~$3.47\%$ across all folds and ``DC pipes'' are within the average with an error of~$8.05\%$, all other three sub-structures have way larger classification error, with~$42.86\%$ of ``UC spheres'' being wrongly classified, see Table~\ref{tab:FoldDistribution}.
This confirms the impression from Figure~\ref{fig:ViolinPlotVolume} that ``UC pipes'' and ``UC spheres'' are not well separated solely by their volume.

\begin{table}
	\centering
	\begin{tabular}{l||r|r|r|r|r||r}
		 			& Fold1 &Fold2 &Fold3 &Fold4 &Fold5 & Error (\%)\\
	 \hline \hline
	``DC spheres'' 	&   116 &  115 &  115 &  115 &  115 & 3.47\\
	\hline
	``DC pipes''	&    77 &   77 &   77 &   77 &   77 &  8.05\\
	\hline
	``areas''		&    29 &   30 &   29 &   30 &   29 &  14.97\\
	\hline
	``UC pipes''	&    20 &   20 &   20 &   20 &   21 &  18.81\\
	\hline
	``UC spheres''	&     6 &    6 &    5 &    6 &    5 &  42.86\\
	\hline \hline
	Error (\%)		& 7.26	& 9.27 & 8.06 & 9.68 & 7.66
	\end{tabular}
	\caption{Distribution of the different sub-structures over the folds of the data set used for cross-correlation. The errors given indicate classification errors of the sub-structures solely based on the volume of the sub-structure.}
	\label{tab:FoldDistribution}
\end{table}

The scatterplots in Figure~\ref{fig:resTopViewAndScatterplots} suggest that the surface area of the sub-structure does not carry a lot of information regarding the type of the structure, but that the activator radius, in conjunction with the volume could provide a better classifier.
On the same folds as used in the training of the above predictor, we train a $k$-nearest neighbor predictor.
The used neighborhood sizes were~${k\in\{1,\ldots,20\}}$.
As the two features, volume and activator radius, exhibit very different sizes, we normalize these two dimensions to exclude any artifacts.
For our data set and the number of chosen folds, we obtain a minimal classification error for~$k=1$, where the mean classification error over all folds is~$7.35\%$.
Thus, most of the information on sub-structures is already captured by the volume and the activator radius only provides a minimal gain in information.

Considering that the data is randomly generated and hand-classified, the predictors presented here already obtain a high level of accuracy.
It is clear from our findings that the volume of a sub-structure is the largest predictor for its type.
Further research is necessary to provide even more precise classifications.

\section{Conclusion}
\label{sec:Conclusion}

In this paper, we have presented a visually-guided investigation of sub-structures in three-dimensional Turing-like patterns.
Based on a discretization, we gave a complete partition of the parameter space and found five different non-trivial sub-structures, see Figure~\ref{fig:3DPatterns}.
Thereby, we provided a positive answer to the conjecture of~\cite{skrodzki2017turing}, who proposed the existence of two-dimensional sub-structures in three-dimensional Turing-like patterns.
Based on our experiments, we furthermore provided statistical insight into the distribution of the different sub-structures within the parameter space.
From these data, we presented several conjectures both regarding three-dimensional and higher-dimensional Turing-like patterns.

The experimental approach chosen in this paper has several weaknesses.
First, the discretization of the activation probability, as given in Equation~(\ref{equ:sigmoid}), is not linear, but emphasizes very low and very high activation probabilities equally.
Therefore, our statistical findings as reported in the histogram in Figure~\ref{fig:StructureHistogram} are correspondingly distorted.
Second, the identification and distinction of the different sub-structure cases depends solely on visual inspection, the current iteration number, and subjective interpretation.
Aside from the clearly distinguishable ``DC all'' and ``UC all'' cases, there is no automatism for classifying the sub-structures yet.

Future research has to be directed at finding a thorough mathematical description of the different sub-structures described in this work.
We have shown that the volume is a strong predictor to classify the ``DC'' sub-structures, but it does not work well on the ``UC'' sub-structures.
Furthermore, a larger portion of the parameter space has to be investigated to better understand the zero-, one-, and two-dimensional sub-structures that arise.
The experiments presented in this paper can help narrow down a corresponding portion of the parameter space and thus help focus on the relevant parts.

%

\section*{Funding}

This material is based upon work supported by the National Science Foundation under Grant No. DMS-1439786 and the Alfred P. Sloan Foundation award G-2019-11406 while the author was in residence at the Institute for Computational and Experimental Research in Mathematics in Providence, RI, during the Illustrating Mathematics program.\\
Furthermore, this research was supported by the DFG Collaborative Research Center TRR 109, ``Discretization in Geometry and Dynamics'', RIKEN iTHEMS, and the German National Academic Foundation.


\bibliographystyle{tfcad}
\bibliography{bibliography}

\begin{thebibliography}{35}
\newcommand{\enquote}[1]{``#1''}
\providecommand{\natexlab}[1]{#1}
\providecommand{\url}[1]{\normalfont{#1}}
\providecommand{\urlprefix}{}

\bibitem[Asai et~al.(1999)]{asai1999zebrafish}
Asai, Rihito, Emiko Taguchi, Yukari Kume, Mayumi Saito, and Shigeru Kondo.
  1999. ``Zebrafish leopard gene as a component of the putative
  reaction-diffusion system.'' \emph{Mechanisms of development} 89 (1-2):
  87--92.

\bibitem[B{\'a}ns{\'a}gi, Vanag, and Epstein(2011)]{bansagi2011tomography}
B{\'a}ns{\'a}gi, Tam{\'a}s, Vladimir~K. Vanag, and Irving~R. Epstein. 2011.
  ``Tomography of reaction-diffusion microemulsions reveals three-dimensional
  Turing patterns.'' \emph{Science} 331 (6022): 1309--1312.

\bibitem[Beer(2004)]{beer2004autopoiesis}
Beer, Randall~D. 2004. ``Autopoiesis and cognition in the game of life.''
  \emph{Artificial Life} 10 (3): 309--326.

\bibitem[Beer(2014)]{beer2014cognitive}
Beer, Randall~D. 2014. ``The cognitive domain of a glider in the game of
  life.'' \emph{Artificial life} 20 (2): 183--206.

\bibitem[Beer(2015)]{beer2015characterizing}
Beer, Randall~D. 2015. ``Characterizing autopoiesis in the game of life.''
  \emph{Artificial Life} 21 (1): 1--19.

\bibitem[Beer(2020)]{beer2020investigation}
Beer, Randall~D. 2020. ``An investigation into the origin of autopoiesis.''
  \emph{Artificial Life} 26 (1): 5--22.

\bibitem[Bressloff et~al.(2002)]{bressloff2002geometric}
Bressloff, Paul~C., Jack~D. Cowan, Martin Golubitsky, Peter~J. Thomas, and
  Matthew~C. Wiener. 2002. ``What geometric visual hallucinations tell us about
  the visual cortex.'' \emph{Neural computation} 14 (3): 473--491.

\bibitem[Chan(2018)]{chan2018lenia}
Chan, Bert Wang-Chak. 2018. ``Lenia-biology of artificial life.'' \emph{arXiv
  preprint arXiv:1812.05433} .

\bibitem[Gardener(1970)]{gardener1970mathematical}
Gardener, Martin. 1970. ``Mathematical Games: The fantastic combinations of
  John Conway's new solitaire game ``life''.'' \emph{Scientific American} 223:
  120--123.

\bibitem[Greenfield(2016)]{greenfield2016turing}
Greenfield, Gary~R. 2016. ``Turing-like Patterns from Cellular Automata.'' In
  \emph{Proceedings of Bridges 2016: Mathematics, Music, Art, Architecture,
  Education, Culture},  edited by Eve Torrence, Bruce Torrence, Carlo~H.
  S\'equin, Douglas McKenna, Krist\'of Fenyvesi, and Reza Sarhangi, Phoenix,
  Arizona, 151--158. Tessellations Publishing. Available online at
  \url{http://archive.bridgesmathart.org/2016/bridges2016-151.html}.

\bibitem[Hiscock and Megason(2015)]{hiscock2015orientation}
Hiscock, Tom~W., and Sean~G. Megason. 2015. ``Orientation of turing-like
  patterns by morphogen gradients and tissue anisotropies.'' \emph{Cell
  systems} 1 (6): 408--416.

\bibitem[Ishida(2020)]{ishida2020emergence}
Ishida, Takeshi. 2020. ``Emergence of Turing Patterns in a Simple Cellular
  Automata-Like Model via Exchange of Integer Values between Adjacent Cells.''
  \emph{Discrete Dynamics in Nature and Society} 2020.

\bibitem[Kondo and Miura(2010)]{kondo2010reaction}
Kondo, Shigeru, and Takashi Miura. 2010. ``Reaction-diffusion model as a
  framework for understanding biological pattern formation.'' \emph{science}
  329 (5999): 1616--1620.

\bibitem[Lepp{\"a}nen et~al.(2002)]{leppanen2002new}
Lepp{\"a}nen, Teemu, Mikko Karttunen, Kimmo Kaski, Rafael~A Barrio, and Limei
  Zhang. 2002. ``A new dimension to Turing patterns.'' \emph{Physica D:
  Nonlinear Phenomena} 168: 35--44.

\bibitem[McCabe(2010)]{mccabe2010cyclic}
McCabe, Jonathan. 2010. ``Cyclic Symmetric Multi-Scale Turing Patterns.'' In
  \emph{Proceedings of Bridges 2010: Mathematics, Music, Art, Architecture,
  Culture},  edited by George~W. Hart and Reza Sarhangi, Phoenix, Arizona,
  387--390. Tessellations Publishing. Available online at
  \url{http://archive.bridgesmathart.org/2010/bridges2010-387.html}.

\bibitem[Merle, Messio, and Mozziconacci(2019)]{merle2019turing}
Merle, M{\'e}lody, Laura Messio, and Julien Mozziconacci. 2019. ``Turing-like
  patterns in an asymmetric dynamic Ising model.'' \emph{Physical Review E} 100
  (4): 042111.

\bibitem[Nakamasu et~al.(2009)]{nakamasu2009interactions}
Nakamasu, Akiko, Go~Takahashi, Akio Kanbe, and Shigeru Kondo. 2009.
  ``Interactions between zebrafish pigment cells responsible for the generation
  of Turing patterns.'' \emph{Proceedings of the National Academy of Sciences}
  106 (21): 8429--8434.

\bibitem[Polthier et~al.(2002)]{polthier2002publication}
Polthier, Konrad, Samy Khadem, Eike Preu{\ss}, and Ulrich Reitebuch. 2002.
  ``Publication of interactive visualizations with Java View.'' In
  \emph{Multimedia tools for communicating mathematics}, 241--264. Springer.

\bibitem[Polthier et~al.(2020)]{polthier2020javaview}
Polthier, Konrad, Ulrich Reitebuch, Sunil~Kumar Yadav, and Eric Zimmermann.
  2020. ``JavaView.'' V.~5.0.1, Accessed 2020-06-09.
  \urlprefix\url{http://javaview.de/}.

\bibitem[Pringle and Tarnita(2017)]{pringle2017spatial}
Pringle, Robert~M., and Corina~E. Tarnita. 2017. ``Spatial self-organization of
  ecosystems: integrating multiple mechanisms of regular-pattern formation.''
  \emph{Annual review of Entomology} 62: 359--377.

\bibitem[Schwehm(2016)]{schwehm2016fast}
Schwehm, Markus. 2016. ``A Fast Algorithm for Creating Turing-McCabe
  Patterns.'' In \emph{Proceedings of Bridges 2016: Mathematics, Music, Art,
  Architecture, Education, Culture},  edited by Eve Torrence, Bruce Torrence,
  Carlo~H. S\'equin, Douglas McKenna, Krist\'of Fenyvesi, and Reza Sarhangi,
  Phoenix, Arizona, 431--434. Tessellations Publishing. Available online at
  \url{http://archive.bridgesmathart.org/2016/bridges2016-431.html}.

\bibitem[Scoones and Hiscock(2020)]{scoones2020dot}
Scoones, Jake~Cornwall, and Tom~W. Hiscock. 2020. ``A dot-stripe Turing model
  of joint patterning in the tetrapod limb.'' \emph{Development} 147 (8).

\bibitem[Skrodzki and Polthier(2017)]{skrodzki2017turing}
Skrodzki, Martin, and Konrad Polthier. 2017. ``Turing-Like Patterns Revisited:
  A Peek Into The Third Dimension.'' In \emph{Proceedings of Bridges 2017:
  Mathematics, Art, Music, Architecture, Education, Culture},  edited by David
  Swart, Carlo~H. S\'equin, and Krist\'of Fenyvesi, Phoenix, Arizona, 415--418.
  Tessellations Publishing. Available online at
  \url{http://archive.bridgesmathart.org/2017/bridges2017-415.pdf}.

\bibitem[Skrodzki and Polthier(2018)]{skrodzki2018mondrian}
Skrodzki, Martin, and Konrad Polthier. 2018. ``Mondrian Revisited: A Peek Into
  The Third Dimension.'' In \emph{Proceedings of Bridges 2018: Mathematics,
  Art, Music, Architecture, Education, Culture},  edited by Eve Torrence, Bruce
  Torrence, Carlo S\'equin, and Krist\'of Fenyvesi, Phoenix, Arizona, 99--106.
  Tessellations Publishing. Available online at
  \url{http://archive.bridgesmathart.org/2018/bridges2018-99.pdf}.

\bibitem[Skrodzki, Reitebuch, and Polthier(2016)]{skrodzki2016chladni}
Skrodzki, Martin, Ulrich Reitebuch, and Konrad Polthier. 2016. ``Chladni
  Figures Revisited: A Peek Into The Third Dimension.'' In \emph{Proceedings of
  Bridges 2016: Mathematics, Music, Art, Architecture, Education, Culture},
  edited by Eve Torrence, Bruce Torrence, Carlo S\'equin, Douglas McKenna,
  Krist\'of Fenyvesi, and Reza Sarhangi, Phoenix, Arizona, 481--484.
  Tessellations Publishing. Available online at
  \url{http://archive.bridgesmathart.org/2016/bridges2016-481.html}.

\bibitem[Swindale(1980)]{swindale1980model}
Swindale, Nicolas~V. 1980. ``A model for the formation of ocular dominance
  stripes.'' \emph{Proceedings of the Royal Society of London. Series B.
  Biological Sciences} 208: 243--264.

\bibitem[Thompson(1917)]{thompson1917growth}
Thompson, d'Arcy~Wentworth. 1917. \emph{On Growth and Form}. Cambridge
  University Press.

\bibitem[Tonello and Siebert(2019)]{tonello2019boolean}
Tonello, Elisa, and Heike Siebert. 2019. ``Boolean analysis of lateral
  inhibition.'' \emph{arXiv preprint arXiv:1904.02544} .

\bibitem[Turing(1952)]{turing1990chemical}
Turing, Alan~Mathison. 1952. ``The chemical basis of morphogenesis.''
  \emph{Philosophical Transactions of the Royal Society of London. Series B,
  Biological Sciences} 237 (641): 37--72.

\bibitem[Weide~Rodrigues, Mistro, and
  D{\'\i}az~Rodrigues(2020)]{weide2020pattern}
Weide~Rodrigues, Vagner, Diomar~Cristina Mistro, and Luiz~Alberto
  D{\'\i}az~Rodrigues. 2020. ``Pattern Formation and Bistability in a
  Generalist Predator-Prey Model.'' \emph{Mathematics} 8 (1): 20.

\bibitem[Werth(2015)]{werth2015turing}
Werth, Andrew. 2015. ``Turing Patterns in Photoshop.'' In \emph{Proceedings of
  Bridges 2015: Mathematics, Music, Art, Architecture, Culture},  edited by
  Kelly Delp, Craig~S. Kaplan, Douglas McKenna, and Reza Sarhangi, Phoenix,
  Arizona, 459--462. Tessellations Publishing. Available online at
  \url{http://archive.bridgesmathart.org/2015/bridges2015-459.html}.

\bibitem[Wertheim and Roose(2019)]{wertheim2019can}
Wertheim, Kenneth~Y, and Tiina Roose. 2019. ``Can VEGFC Form Turing Patterns in
  the Zebrafish Embryo?'' \emph{Bulletin of mathematical biology} 81 (4):
  1201--1237.

\bibitem[Wolfram(2002)]{wolfram2002new}
Wolfram, Stephen. 2002. \emph{A new kind of science}. Wolfram media Champaign,
  IL.

\bibitem[Woolley, Baker, and Maini(2017)]{woolley2017turing}
Woolley, Thomas~E., Ruth~E. Baker, and Philip~K. Maini. 2017. ``Turing's theory
  of morphogenesis.'' In \emph{The Turing Guide},  edited by B.~Jack Copeland,
  Jonathan~P. Bowen, Mark Sprevak, and Robin Wilson, 201--213. Oxford
  University Press. An optional note.

\bibitem[Young(1984)]{young1984local}
Young, David~A. 1984. ``A local activator-inhibitor model of vertebrate skin
  patterns.'' \emph{Mathematical Biosciences} 72 (1): 51--58.

\end{thebibliography}

\end{document}